\documentclass{aa}
\usepackage{txfonts}
\usepackage{graphicx}
\usepackage{natbib}
\bibpunct{(}{)}{;}{a}{}{,}
\begin{document}

\title{Observations and asteroseismic analysis of the rapidly pulsating hot B subdwarf PG 0911+456\thanks{This study made extensive use of the computing facilities offered by the Calcul en Midi-Pyr\'en\'ees (CALMIP) project and the Centre Informatique National de l'Enseignement Sup\'erieur (CINES), France. Some of the spectroscopic observations reported here were obtained at the MMT Observatory, a joint facility of the University of Arizona and the Smithsonian Institution.}}
\author{
S.K. Randall \inst{1}
\and E.M. Green \inst{2}
\and V. Van Grootel \inst{3,4}
\and G. Fontaine \inst{3}
\and S. Charpinet \inst{4}
\and M. Lesser \inst{2}
\and P. Brassard \inst{3}
\and T. Sugimoto \inst{2}
\and P. Chayer \inst{5,6}
\and A. Fay \inst{2}
\and P. Wroblewski \inst{2}
\and M. Daniel \inst{2}
\and S. Story \inst{2}
\and T. Fitzgerald \inst{2}
}

\institute{
ESO, Karl-Schwarzschild-Str. 2, 85748 Garching bei M\"unchen, Germany; \email{srandall@eso.org}
\and Steward Observatory, University of Arizona, 933 North Cherry Avenue, Tucson, AZ 85721, USA
\and Universit\'e de Montr\'eal, C.P. 6128, Succ. Centre-Ville, Montr\'eal, QC H3C 3J7, Canada
\and UMR 5572, Universit\'e Paul Sabatier et CNRS, Observatoire Midi-Pyr\'en\'ees, 14 Av. E. Belin, 31400, Toulouse, France
\and Department of Physics and Astronomy, John Hopkins University, 3400 North Charles Street, Baltimore, MD 21218-2686, USA
\and Primary affiliation: Department of Physics and Astronomy, University of Victoria, PO Box 3055, Victoria, BC V8W 3P6, Canada
}

\date{Received date / Accepted date}

\abstract
{}
{The principal aim of this project is to determine the structural parameters of the rapidly pulsating subdwarf B star PG 0911+456 from asteroseismology. Our work forms part of an ongoing programme to constrain the internal characteristics of hot B subdwarfs with the long-term goal of differentiating between the various formation scenarios proposed for these objects. So far, a detailed asteroseismic interpretation has been carried out for 6 such pulsators, with apparent success. First comparisons with evolutionary theory look promising, however it is clear that more targets are needed for meaningful statistics to be derived.}
{The observational pulsation periods of PG 0911+456 were extracted from rapid time-series photometry using standard Fourier analysis techniques. Supplemented by spectroscopic estimates of the star's mean atmospheric parameters, they were used as a basis for the "forward modelling" approach in asteroseismology. The latter culminates in the identification of one or more "optimal" models that can accurately reproduce the observed period spectrum. This naturally leads to an identification of the oscillations detected in terms of degree $\ell$ and radial order $k$, and infers the structural parameters of the target.}
{The high S/N low- and medium resolution spectroscopy obtained led to a refinement of the atmospheric parameters for PG 0911+456, the derived values being $T_{\rm eff}$=31,940$\pm$220 K, $\log{g}$=5.767$\pm$0.029, and $\log{\rm He/H}$=$-$2.548$\pm$0.058. From the photometry it was possible to  extract 7 independent pulsation periods in the 150$-$200 s range with amplitudes between 0.05 and 0.8 \% of the star's mean brightness. There was no indication of fine frequency splitting over the $\sim$68-day time baseline, suggesting a very slow rotation rate. An asteroseismic search of parameter space identified several models that matched the observed properties of PG 0911+456 well, one of which was isolated as the "optimal" model on the basis of spectroscopic and mode identification considerations. All the observed pulsations are identified with low-order acoustic modes with degree indices $\ell$=0,1,2 and 4, and match the computed periods with a dispersion of only $\sim$ 0.26 \%, typical of the asteroseismological studies carried out to date for this type of star. The inferred structural parameters of PG 0911+456 are $T_{\rm eff}$=31,940$\pm$220 K (from spectroscopy) , $\log{g}$=5.777$\pm$0.002, $M_{\ast}/M_{\odot}$=0.39$\pm$0.01, $\log{M_{env}/M_{\ast}}$=$-$4.69$\pm$0.07, $R/R_{\odot}$=0.133$\pm$0.001 and $L/L_{\odot}$=16.4$\pm$0.8. We also derive the absolute magnitude $M_V$=4.82$\pm$0.04 and a distance $d$=930.3$\pm$27.4 pc.}
{}

\keywords{asteroseismology, subdwarfs, pulsating stars}
\titlerunning{Asteroseismic analysis of PG 0911+456}
\maketitle
\authorrunning{S.K. Randall et al.}

\section{INTRODUCTION}

Subdwarf B (sdB) stars are evolved extreme horizontal branch stars with atmospheric parameters in the 20,000 K $\lesssim T_{\rm eff} \lesssim$ 40,000 K and  5.0 $\lesssim \log{g} \lesssim$ 6.2 range \citep{heber1986}. They are believed to be composed of helium-burning cores surrounded by thin hydrogen-rich envelopes and are characterised by a narrow mass distribution strongly peaked at $\sim$ 0.48 $M_{\odot}$ \citep{dorman1993}.  While it is generally accepted that they evolved from the red giant branch (RGB) and constitute the immediate progenitors of low-mass white dwarfs \citep{bergeron1994}, the details of their formation are not yet understood. It does however seem clear that sdB progenitors lost a significant fraction of their envelope mass near the tip of the first RGB, leaving them with insufficient fuel to ascend the asymptotic giant branch (AGB) after core helium exhaustion. Plausible formation channels were modelled in detail by, e.g., \citet{han2002,han2003} and include binary evolution via a common envelope (CE) phase, stable Roche lobe overflow (RLOF), and the merger of two helium white dwarfs. These distinct evolutionary scenarios should leave a clear imprint not only on the binary distribution of sdB stars (CE evolution will produce sdB's in very close binary systems, RLOF gives rise to much longer period binaries, and the white dwarf merger results in single sdB stars), but also on their mass and hydrogen-envelope thickness distribution. Observational surveys focusing on radial velocity variations and the spectroscopic detection of companions have recently established that at least half of the sdB population reside in binaries (e.g. \citet{allard1994,ulla1998}), a significant fraction of them having short orbital periods from hours to days (\citet{green1997,maxted2001}). Accurate determinations of the stars' internal parameters on the other hand are harder to come by using traditional techniques; the mass has so far been measured only for the very rare case of an eclipsing binary \citep{wood1999}, while the envelope thickness eludes direct study.

Fortunately, a small fraction ($\sim$ 5 $\%$) of sdB stars have been discovered to exhibit rapid, multi-periodic luminosity variations on a time-scale of hundreds of seconds, thus opening up the possibility of using asteroseismology to constrain their internal parameters. Since the near-simultaneous theoretical prediction \citep{charp1996, charp1997} and observational discovery \citep{kilkenny1997} of the so-called EC 14026 stars, both the modelling and measurement of their pulsational properties have come a long way (see \citet{fontaine2006} for a review). Simulating the pulsation spectra of a large grid of sdB models in terms of low-degree, low-order $p$-modes and numerically determining the "optimal" model that best fits a series of observed periodicities has so far resulted in the asteroseismological determination of the internal parameters for six EC 14026 pulsators: PG 0014+067 \citep{brassard2001}, PG 1047+003 \citep{charp2003}, PG 1219+534 \citep{charp2005a}, Feige 48 \citep{charp2005b}, EC 20117-4014 \citep{randall2006c}, and PG 1325+101\citep{charp2006}. These first asteroseismic results show promising trends as far as matching the expected mass and hydrogen shell thickness distribution is concerned, however more targets are needed to start assessing the importance and accuracy of the proposed formation channels. Here we present an asteroseismological analysis of the subdwarf B star PG 0911+456 based on photometry obtained with the new Mont4kccd at Mt. Bigelow, Arizona. In the next sections we describe the observations and frequency analysis, followed by the asteroseismic exploitation of the target and a discussion of the internal parameters inferred.

\section{OBSERVATIONS}

\subsection{Spectroscopy $\&$ atmospheric analysis}

PG 0911+456 was first identified as a hot star in the Palomar-Green survey \citep{green1986}, and analysed in more detail by \citet{koen1999} using a medium-resolution ($\sim$0.8\AA/pix) spectrum obtained on the 4.2-m William Herschel telescope on La Palma, Spain. Fitting a pure hydrogen line profile grid to the data suggested atmospheric parameters of $T_{\rm eff}$=31,900$\pm$200 K and $\log{g}$=5.82$\pm$0.02 , while an independent analysis carried out by Rex Saffer produced consistent values of $T_{\rm eff}$=31,400 K and $\log{g}$=5.80. This is typical for an EC 14026 star, and places PG 0911+456 in the central part of the theoretical instability strip \citep{charp2001}. The authors also noted that the helium lines were weak for this object, and that there was no obvious spectroscopic signature of a companion. 

\begin{figure}[t]
\resizebox{\hsize}{!}{\includegraphics{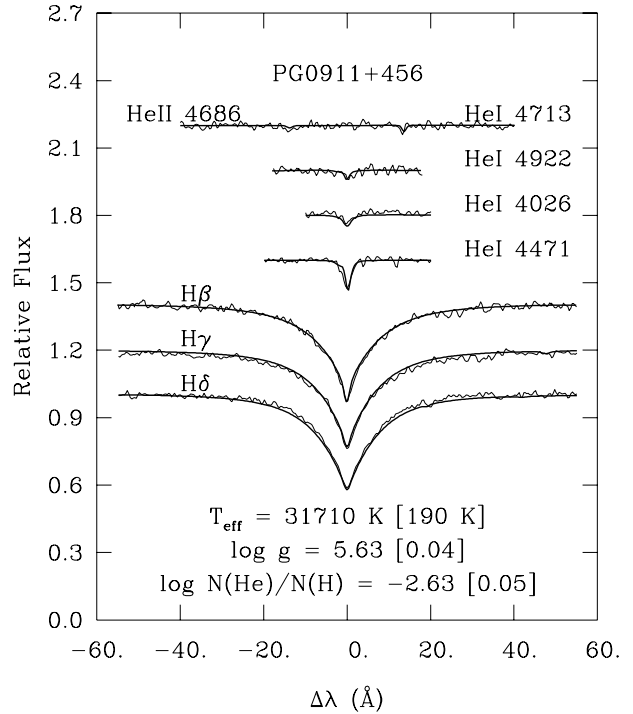}}
\caption{Model fit (solid curve) to the hydrogen Balmer lines and helium lines available in our time averaged high S/N ratio, medium-resolution optical spectrum of PG 0911+456.}
\label{medres}
\end{figure}

\begin{figure}[t]
\resizebox{\hsize}{!}{\includegraphics{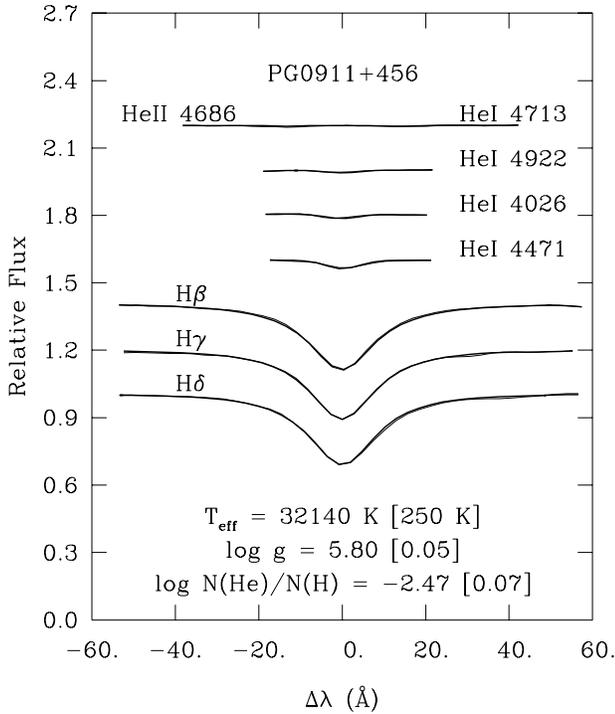}}
\caption{Model fit (solid curve) to the hydrogen Balmer lines and helium lines available in our time averaged high S/N ratio, low-resolution optical spectrum of PG 0911+456.}
\label{lowres}
\end{figure}

We obtained additional spectra with Steward Observatory's 2.3-m telescope on Kitt Peak, Arizona, as well as the 6.5-m MMT as part of an ongoing program to determine the atmospheric characteristics of a large sample of both pulsating and non-pulsating sdB stars in a homogeneous way (for details see Green, Fontaine, $\&$ Chayer, in preparation). The Kitt Peak data cover the $\sim$3600$-$5240 $\AA$ range at low wavelength resolution ($\sim$9 $\AA$), while the MMT spectrum concentrates on the $\sim$4000$-$4950 $\AA$ region at a higher resolution of $\sim$1 $\AA$. Both data sets were analysed with the help of a detailed grid of non-LTE model atmospheres and synthetic spectra designed especially for subdwarf B stars. These were computed using the public codes TLUSTY and SYNSPEC \citep{hubeny1995,lanz1995}, and in contrast to the models employed by \citet{koen1999} incorporate helium but no metals. We show the best model fit to the available Balmer and neutral helium lines for the MMT and the 2.3-m spectra in Figures \ref{medres} and \ref{lowres} respectively, and also indicate the atmospheric parameters inferred. Note that the values in brackets refer to the formal fitting errors. While both the effective temperature and surface gravity derived from the low-resolution spectrum agree well with those reported by \citet{koen1999}, the value of $\log{g}$ from the higher resolution data is significantly lower. This is somewhat surprising, since in our experience the atmospheric parameters computed from both types of spectra are compatible as long as the S/N of the data is high enough. Examining the figures, we find the fits to both spectra to be satisfactory, however it is apparent that the MMT data are far noisier than those from Kitt Peak. Indeed, comparing the quality of the single medium-resolution spectrum obtained for PG 0911+456 to equivalent data for previously analysed subdwarf B stars such as PG 1325+101 \citep{charp2006} or PG 1219+534 \citep{charp2005a} reveals a significantly lower S/N ($\sim$78 compared to typically 200+) for the case of our target. In contrast, our combined low-resolution spectrum of PG 0911+456 is of outstanding quality (S/N$\sim$350). Given that the atmospheric parameters derived from the latter are also in excellent agreement with those reported by \citet{koen1999}, we believe them to be more accurate than those obtained from the MMT spectra.  This was also the conclusion reached by the referee (U. Heber), who kindly re-analysed our Kitt Peak spectrum for PG 0911+456 using independent models that incorporate metals at both solar and sub-solar abundances, and derived atmospheric parameters consistent with our low-resolution estimates. He also noted that, in his analysis of MMT medium-resolution spectra of about a dozen sdB stars, consistently lower surface gravities and temperatures were found than from  their low-resolution counterparts, the latter being more reliable due to the wider wavelength coverage and higher number of Balmer lines sampled. 

Using the available estimates of the atmospheric parameters for PG 0911+456 we compute a weighted mean of $T_{\rm eff}$=31,940$\pm$220 K and $\log{g}$=5.767$\pm$0.029. As already noted by \citet{koen1999}, the atmospheric helium abundance is low even for a subdwarf B star at $\log{\rm He/H}$=$-$2.548$\pm$0.058.

\subsection{Time-series photometry $\&$ frequency analysis}

PG 0911+456 was first identified as a member of the (then) newly discovered class of rapidly pulsating EC 14026 stars by \citet{koen1999}. On the basis of $\sim$17 hours of Str\"omgen $b$ band time-series photometry gathered at the 0.9-m McDonald Observatory on Mt. Locke, Texas, the authors were able to extract 3 convincing periodicities between 150 s and 170 s (see Table 2 below) and estimate an approximate mean Str\"omgen $b$ magnitude of $\sim$14.6.  

Our follow-up photometry was obtained with the new Mont4kccd (also known as La Poune II) on Steward Observatory's 1.55-m Mt. Bigelow telescope in Arizona. Since this constitutes the first published science data obtained with the instrument, we give a brief overview of its characteristics and the observing strategy developed for fast time-series photometry. The 4k$\times$4k CCD is a Fairchild CCD 486 detector with 4096$\times$4097 15 micron pixels that has been processed for backside illumination at the Universtity of Arizona Imaging Technology Laboratory (ITL). Treatment with the ITL Chemiabsorption Charging and antireflection coating results in a good sensitivity between $\sim$3000 and 8000 $\AA$, with a quantum efficiency peak of about 95$\%$ at 4500 $\AA$. The read noise of the detector is 5.0 e$^-$ (at a readout rate of about 50 kHz per channel), the dark current is 16.6 e$^-$/pixel/hour at $-$130$^{\circ}$ C, the unbinned full well capacity is 131,000 e$^-$ and the device is operated at a gain of approximately 3.1 e$^-$/DN. It is packaged in a Kovar tub and is steered with an Astronomical Research Cameras, Inc.\ Gen2 controller using two video channels. Target acquisition is performed with the help of the AzCam software developed by ITL and Steward Observatory. The Mont4kccd is equipped with a suite of filters (Bessell U, Harris BVR, Arizona I, as well as the broadband WFPC2 F555W and F814 and the Schott 8612 filters, among others \footnote{The Mont4kccd QE curve as well as the filter transmission curves can be found on the following webpage: http://james.as.arizona.edu/~psmith/61inch/instruments.html}). Mounted at the Cassegrain focus of the Mt. Bigelow telescope, the pixel scale is 0.14"/pixel, yielding a field of view of 9.7'$\times$9.7'. 

The instrument was designed with differential photometry in mind, however the variety of filters available coupled with the excellent sensitivity of the chip and the large field of view make it ideal for many imaging projects. For the purpose of our observations, the main challenge was to reduce the dead-time between exposures to an acceptable level. On-chip binning, normally 3$\times$3 pixels but occasionally 2$\times$2, is ordinarily used to optimise the pixel size on the sky, improve readout speed, and save disk space. However, even 3$\times$3 binning results in an overhead time of 18.2 s, rather long when aiming to temporally resolve the multiple $\sim$150$-$200 s periodicities expected for an EC 14026 pulsator such as PG 0911+456. Our solution was to use 4$\times$4 pixel binning (0.6''/pixel) and read out only the middle 60$\%$ of the rows, reducing the overhead time to just under 10 s. For isolated stars we have in the past been able to obtain relative photometry accurate to 0.001$-$0.002 mags from aperture photometry of images with FWHM's as small as 1.75 pixels. Given that the seeing was never better than $\sim$1'' during our observations of PG 0911+456, this worked quite well for the observations presented here. The most difficult aspect of using a slightly degraded spatial resolution with the Mont4kccd is maintaining the same effective focus throughout the night, as the latter is mainly determined by instrumental flexure related to the pointing position rather than dome temperature. Since keeping the focus constant is necessary to avoid variations in the light curve, we developed a script to monitor the focus of the incoming images.

\begin{table}[h]
\caption{Photometry obtained for PG 0911+456 (2006-2007)}
\label{obslog}
\centering
\begin{tabular}{c c c}
\hline\hline
Date (UT) & Start time (UT) & Length (h) \\
\hline 
20 Dec & 08:10 & 05:07 \\
21 Dec & 07:40 & 05:37 \\
15 Jan & 06:04 & 07:21 \\
16 Jan & 05:30 & 07:54 \\
17 Jan & 04:08 & 09:19 \\
05 Feb & 02:55 & 09:01 \\
10 Feb & 06:11 & 04:05 \\
25 Feb & 02:37 & 08:40 \\
\hline
\end{tabular}
\end{table}

\begin{figure}
\resizebox{\hsize}{!}{\includegraphics{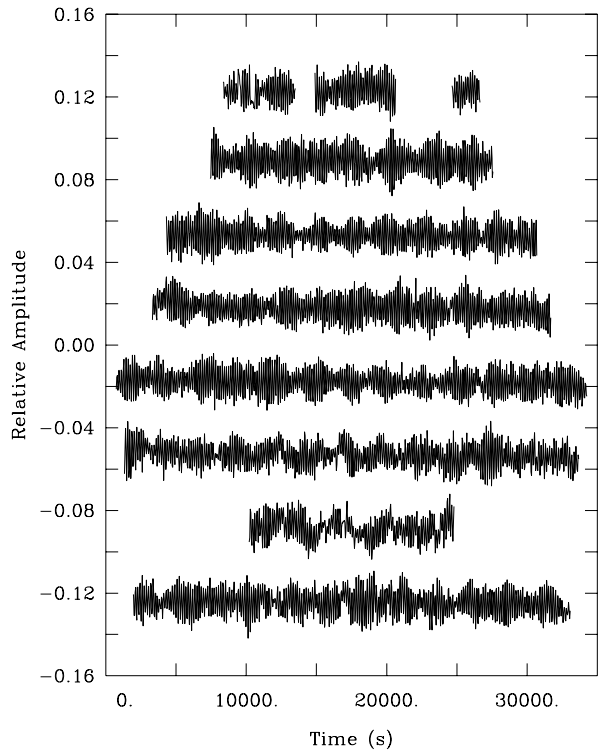}}
\caption{All light curves obtained for PG 0911+456 on the Mt. Bigelow 1.55-m telescope. The data have been shifted arbitrarily along the x and y axes for visualisation purposes. From top to bottom, the curves refer to the nights of 20 Dec, 21 Dec (2006), 15 Jan, 16 Jan, 17 Jan, 05 Feb, 10 Feb and 25 Feb (2007). For details see Table 1.}
\label{allcurves}
\end{figure}

\begin{figure}
\resizebox{\hsize}{!}{\includegraphics[angle=270]{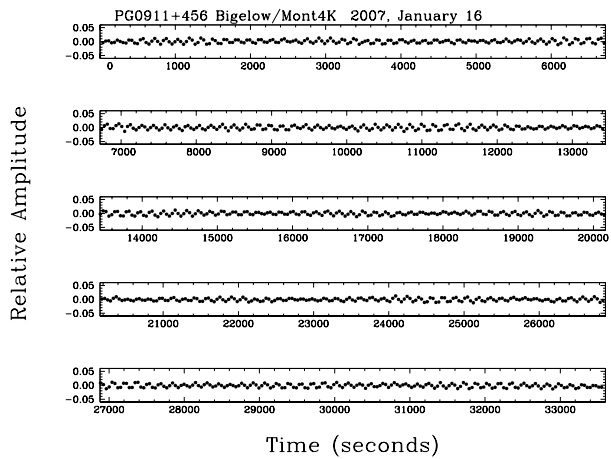}}
\caption{Expanded view of the longest light curve obtained on 16 Jan 2007 in units of fractional brightness intensity and seconds.}
\label{nicecurve}
\end{figure}

We obtained a total of 57 hours of time-series photometry for PG 0911+456 on 8 nights spread over a time baseline of 68 days (see Table \ref{obslog} for details). Owing to the relative faintness of the target, the exposure time was set to 30 s throughout the campaign, and we employed the wide-band Schott 8612 filter to maximise throughput. The data were reduced using standard IRAF reduction aperture photometry routines, except that we set the photometric aperture size separately for each frame to 2.5 times the FWHM in that image. We computed differential light curves of PG 0911+456 on the basis of 6 suitable comparison stars distributed as symmetrically as possible around the target. Figure \ref{allcurves} shows all the nightly light curves obtained; the beating between different modes is clearly visible. However, the remarkable quality of the photometry is better appreciated from the expanded view of a single night's light curve as displayed in Figure \ref{nicecurve}. The relative noise level is reminiscent of that achieved on much larger telescopes, such as the 3.6-m CFHT or the 3.5-m NTT (see, e.g. Figure 2 of \citet{brassard2001}, Figure 4 of \citet{charp2005a} and Figure 2 of \citet{billeres2005}) when taking into account the differences in target brightness. We attribute this partly to the excellent sensitivity of the CCD and our optimised data pipeline, but even more importantly to the relatively high S/N achieved in each exposure and to the excellent spatial distribution of comparison stars around the target. Note that a high data quality can be achieved under less than perfect atmospheric conditions including transparency variations, thin cirrus and a bright Moon, rendering this technique very robust.

\begin{figure}
\resizebox{\hsize}{!}{\includegraphics{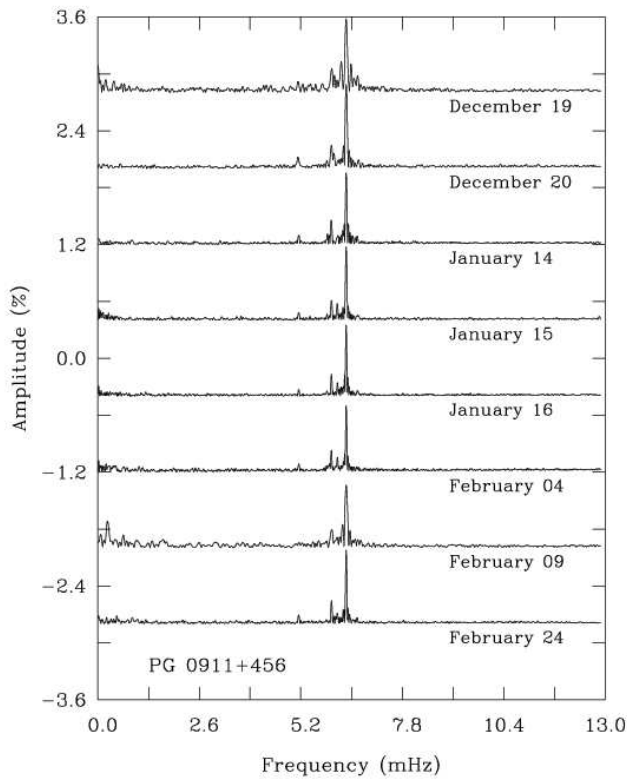}}
\caption{Montage of the nightly Fourier transforms, shifted arbitrarily along the y-axis for visualisation purposes.} 
\label{allft}
\end{figure}

The time-series photometry gathered for PG 0911+456 was analysed in a standard way using a combination of Fourier analysis, least-square fits to the light curve, and pre-whitening techniques (see e.g. \citet{billeres2000} for more details). Figure \ref{allft} shows the Fourier spectra of the individual nightly light curves, which all clearly show a dominant peak at $\sim$ 6.4 mHz as well as a few lower amplitude frequencies in the 6-7 mHz range and an apparently isolated weak oscillation at $\sim$ 5.2 mHz. The power spectrum beyond 13 mHz out to the Nyquist frequency of 26 mHz is entirely consistent with noise and hence not illustrated. Given the poor duty cycle of the data over the long time baseline (3.5 $\%$), it was not obvious that combining all the light curves to give a total Fourier transform would be fruitful and produce reliable peaks (rather than windowing artefacts). We therefore performed a Fourier analysis for both the full dataset and a subset of three consecutive nights (14$-$16 January) and compared the results. Somewhat gratifyingly, the frequencies and amplitudes extracted were the same within the measurement uncertainties and the noise level decreased slightly when adding extra data sets. This has not always been the case for past observational campaigns of EC 14026 stars, where combining light curves well separated in time or of inhomogeneous quality significantly degraded the pulsation spectrum or produced incoherent oscillation peaks (see, e.g. \citet{stobie1997,charp2005a,charp2005b}). The successful merging of different light curves for PG 0911+456 was most likely possible due to the high S/N of the individual light curves as well as a relative homogeneity between the data sets. In what follows, we will focus on the results obtained from all the data, since they correspond to the highest frequency resolution.

\begin{figure}
\resizebox{\hsize}{!}{\includegraphics[bb= 4 106 570 705,clip]{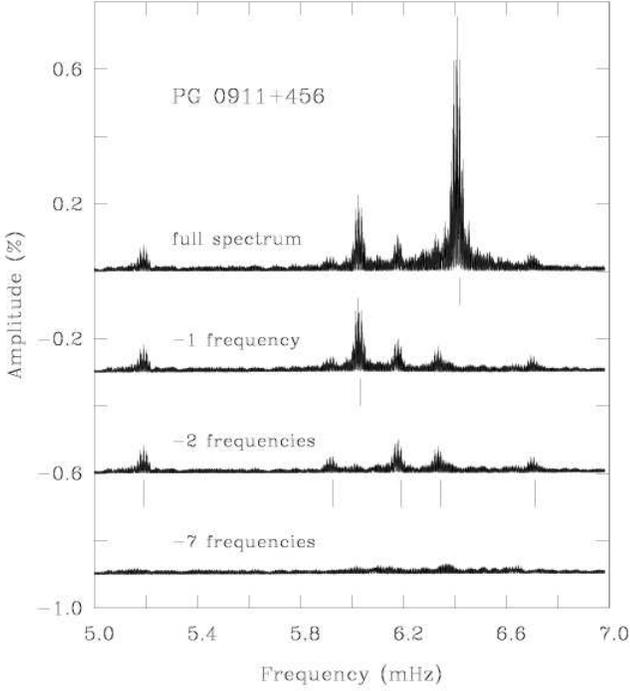}}
\caption{Fourier transform of the entire data set in the 5-7 mHz range (the spectrum outside this range is consistent with noise). The lower transforms show the successive steps of pre-whitening by the dominant, the two strongest and finally all 7 frequencies detected above 4$\sigma$.}
\label{combinedft}
\end{figure}

\begin{figure}
\resizebox{\hsize}{!}{\includegraphics[angle=270]{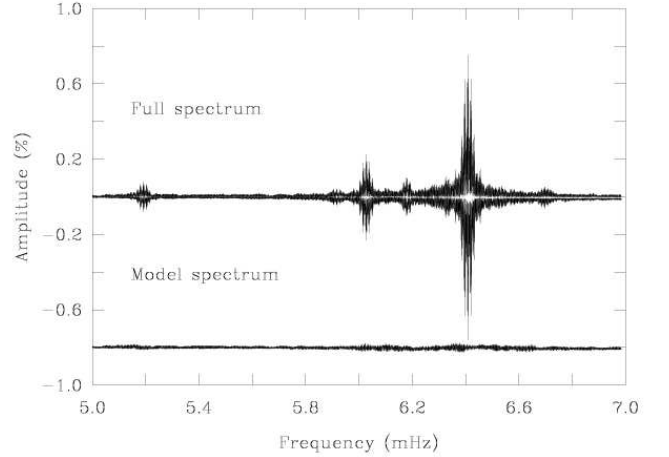}}
\caption{Comparison between the Fourier transform of the entire data set (Full spectrum), and that reconstructed on the basis of the 7 periods, amplitudes and phases derived from least-squares fitting to the light curves (Model spectrum). The lower curve shows the point-by-point difference between the actual FT and the model FT.}
\label{modelspec}
\end{figure}

The Fourier transform of the combined light curve is displayed at the top of Figure \ref{combinedft}, followed by successively pre-whitened spectra in the lower part of the plot. In each case, the number of periodicities extracted is indicated above the spectrum. We detected a total of 7 oscillations down to a threshold of 4 times the noise level in the frequency range plotted (0.07 $\%$). The quality of the pre-whitening can be assessed from Figure \ref{modelspec}, which shows the original Fourier transform together with a noise-less spectrum reconstructed on the basis of the 7 extracted periodicities and the point-by-point difference between the two. While the latter indicates the possible presence of additional, unidentified frequencies, we feel that we have reached the detection limit for convincing oscillations. The periods, frequencies, and amplitudes of the pulsations extracted from the data are listed in Table \ref{frequencies}. It is striking how well the three dominant peaks correspond to the oscillations detected by \citet{koen1999} ($f_{\rm Koen}$) some 9 years previously. Even the amplitude rank order has remained the same, although the two weaker pulsations seem to have diminished somewhat with respect to the main peak (note that the absolute values of the amplitudes cannot be readily compared since the bandpass and frequency resolution of the \citet{koen1999} data are different to our own). Such amplitude variations are by no means uncommon in EC 14026 stars, and can be readily interpreted either in terms of the beating of unresolved modes (e.g. very closely spaced components of a rotationally split multiplet) or significant changes in the intrinsic amplitudes of the modes. In this case, we lean towards the latter explanation since it is likely that at least one of the dominant peaks is associated with a radial mode and therefore not subject to unresolved rotational splitting. And indeed, the two highest amplitude pulsations are identified with $\ell$=0 modes in the asteroseismological analysis presented below. 

\begin{table}
\caption{Oscillations extracted from the combined light curve of PG 0911+456. The uncertainty on the frequencies is estimated to be about 1/10th of the frequency resolution, i.e. around 0.02 $\mu$Hz, while the error on the amplitude is derived from the least-squares fit to the light curve. The frequencies and amplitudes detected by \citet{koen1999} are also listed.}
\label{frequencies}
\centering
\begin{tabular}{cccccc}
\hline\hline
Rank & Period & Frequency & Amplitude &  $f_{\rm Koen}$ & $A_{\rm Koen}$ \\
 & (s) & ($\mu$Hz) & (\%) & ($\mu$Hz) & (\%) \\
\hline
4 & 192.551 & 5193.42 & 0.082$\pm$0.005 & & \\
6 & 168.784 & 5924.73 & 0.054$\pm$0.005 & &\\
2 & 165.687 & 6035.49 & 0.223$\pm$0.005 & 6036.2 & 0.44 \\
3 & 161.554 & 6189.89 & 0.098$\pm$0.005 & 6190.9 & 0.24 \\
5 & 157.581 & 6345.92 & 0.080$\pm$0.005 & & \\
1 & 155.767 & 6419.85 & 0.755$\pm$0.005 & 6419.3 & 0.61 \\
7 & 149.027 & 6710.20 & 0.050$\pm$0.005 & & \\
\hline
\end{tabular}
\end{table}

The absence of frequency splitting over the campaign baseline of over two months implies that PG 0911+456 is a slow rotator with a rotation period $P_{\rm rot}\gtrsim$ 68 d. It is of course also possible that the components of any multiplets have amplitudes below the detection limit, however this seems unlikely given the low noise level of the data. The inferred long rotation period coupled with no sign of a companion being detected from either the time-averaged spectroscopy or the 2MASS $J-H$ colour suggests that our target is a single star. This hypothesis is supported by radial velocity measurements obtained at Steward Observatory's 2.3-m Bok telescope on Kitt Peak as part of an ongoing spectroscopic survey of sdB stars (Green et al, in preparation). No detectable velocity variation was found from ten spectra of PG 0911+456 taken at irregular intervals over a period of three years (2002$-$2005). While at least half of all subdwarf B stars appear to reside in binary systems, these are thought to mostly have short orbital periods (hours to days) and white dwarf secondaries \citep{green1997,maxted2001} or rather long periods (hundreds to thousands of days) and main sequence companions \citep{saffer2001}. Assuming tidally locked rotation, the low noise level of our photometry should reveal rotational frequency splitting in the former case, and a spectroscopic signature of a cooler companion in the latter. Depending on the inclination of the system, the binary motion of the subdwarf may also manifest itself in a Doppler shift of the pulsation frequencies on the time-scale of the orbital period (as detected for the long-period binary EC 20117-4014, see \citet{randall2006c}), of which a comparison of our data with those obtained by \citet{koen1999} show no indication.

\section{Asteroseismic Analysis}

Our asteroseismological study of PG 0911+456 was carried out using the well-known forward method, described in detail by e.g. \citet{brassard2001,charp2005a}. It consists of a double-optimisation procedure that first determines and quantifies the best match between the set of observed periodicities and those calculated for a given model, and subsequently searches parameter space for the model (or models) that can reproduce the data most accurately. The solution is referred to as the "optimal model", and corresponds to the region of parameter space where the goodness-of-fit merit function $S^2$ reaches a minimum. Note that $S^2$ is not a standard $\chi^2$ since no weighting is applied, and is given by
\begin{equation}
S^2=\sum^n_{i=1}(P^i_{obs}-P^i_{theo})^2
\end{equation}
where $P^i_{obs}$ is one of the $n$ periodicities observed and $P^i_{theo}$ is the theoretical period that matches it best.

Over the years we have developed various sets of numerical tools that allow us to apply this scheme to isolate best period fitting models of subdwarf B stars for asteroseismology. Two independent packages, fortunately leading to the same results, are now operational, one in Montr\'eal and the other in Toulouse. Both are built on the simple requirement of \it{global} \rm optimisation, i.e. models aiming to provide a good asteroseismological fit must be able to match all the observed periods simultaneously. Mode identification is normally not assumed a priori, but instead appears naturally as the solution of the global fitting procedure. This provides a convenient cross-check of the "optimal" model's robustness in the case where independent constraints on mode identification are available, e.g. from rotational splitting or multi-colour photometry (see e.g. \citet{charp2005b}, Brassard et al. (2007) in preparation).

The theoretical period spectra are computed on the basis of our so-called "second-generation" models (see \citet{charp1997,charp2001}). These models are static envelope structures incorporating microscopic diffusion under the assumption of an equilibrium having been established between gravitational settling and radiative levitation. Since it is diffusion that creates the iron reservoir responsible $-$ through the $\kappa$-mechanism $-$ for pulsational instabilities in sdB stars, its detailed treatment is essential when predicting the excitation of modes. Moreover, microscopic diffusion changes the stellar structure sufficiently to modify the pulsation periods themselves, and is therefore a necessary ingredient even for the calculation of adiabatic periods. The internal structure of the models is specified by four fundamental input parameters: the effective temperature $T_{\rm eff}$, the surface gravity $\log{g}$, the total stellar mass $M_{\ast}$ and the fractional thickness of the hydrogen-rich envelope $\log{q(\rm H)}=\log{[M(\rm H)/M_{\ast}]}$. The latter parameter is intimately related to the more commonly used quantity $M_{env}$, which corresponds to the total mass of the H-rich envelope\footnote{Note that the parameter $M_{env}$ commonly used in extreme horizontal branch stellar evolution includes the mass of hydrogen contained in the thin He/H transition zone, whereas the parameter $M(H)$ used in our envelope models does not. They can be related with $\log{[M_{env}/M_{\ast}]}=\log{[M(H)/M_{\ast}]}+C$, where $C$ is a small positive term slightly dependent on the model parameters that can be computed from the converged model using the mass of hydrogen present in the transition zone itself.}. Both adiabatic and non-adiabatic oscillation modes are computed for each model using two efficient codes based on finite element techniques. The first solves the four adiabatic oscillation equations and constitutes an intermediate but necessary step to derive estimates for the periods and eigenfunctions that are then used as first guesses for the solution of the full, non-adiabatic set of oscillation equations. The latter computes a number of quantities for each mode with given degree $\ell$ and radial order $k$, most importantly those that can be directly compared with observed quantities, such as the period $P_{th}$ and the stability coefficient $\sigma_I$. If $\sigma_I$ is negative, the mode is stable, while if it is positive the mode is excited, and may therefore be observable if its amplitude is large enough. This can be used as a cross-check for the reliability of the optimal model inferred: obviously, all the periodicities detected should be associated with unstable theoretical modes. Other quantities output by the non-adiabatic code include the kinetic energy $E$ and the rotation coefficient $C_{kl}$ (see Table \ref{periodfit}). The former is of interest mainly for theoretical studies, e.g. when investigating mode trapping, and the latter becomes important only when rotational splitting is detected; therefore they are of secondary interest for our purposes.  

\begin{figure*}
\begin{tabular}{cc}
{\includegraphics[width=8.8cm]{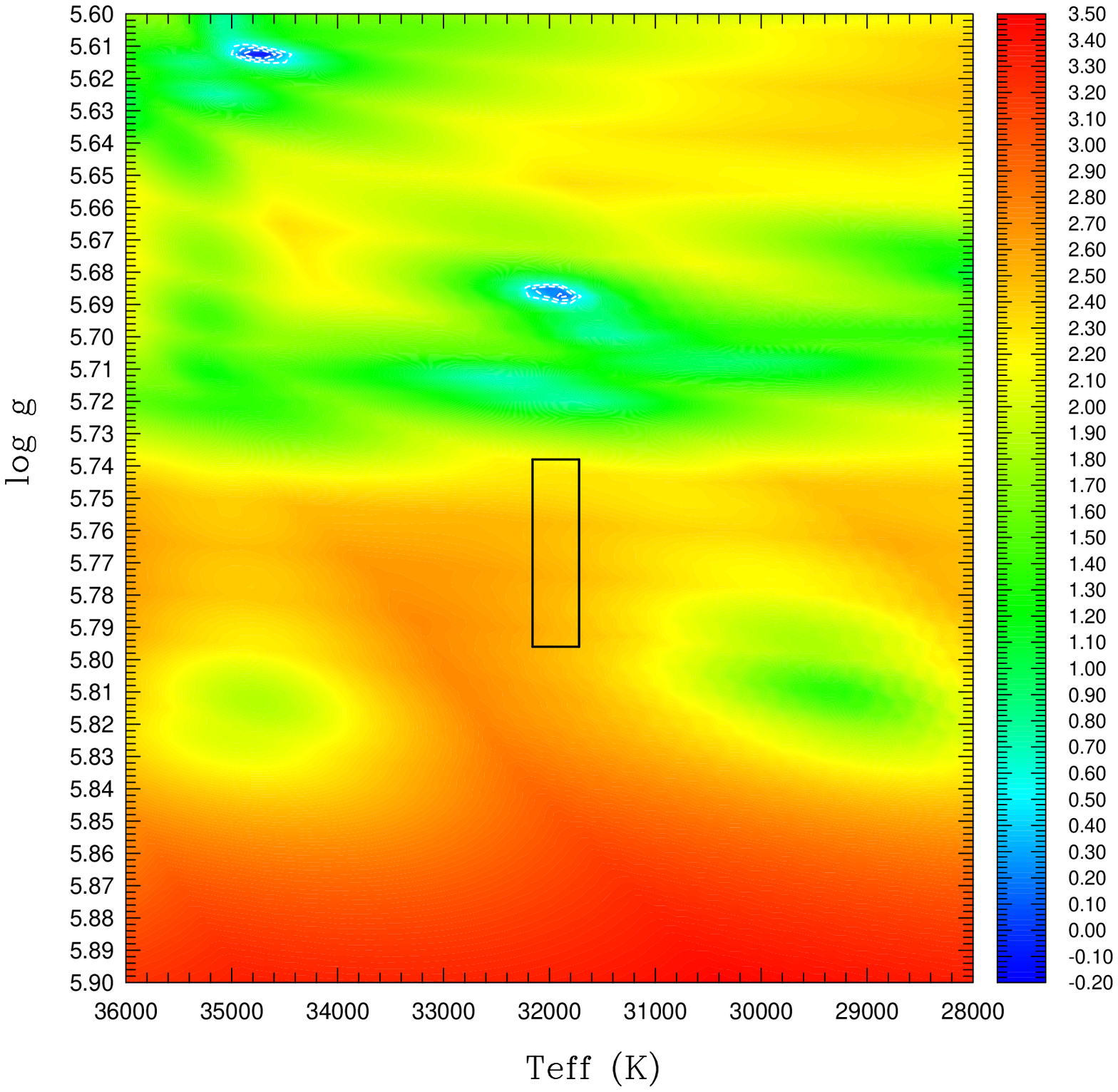}} & {\includegraphics[width=8.8cm]{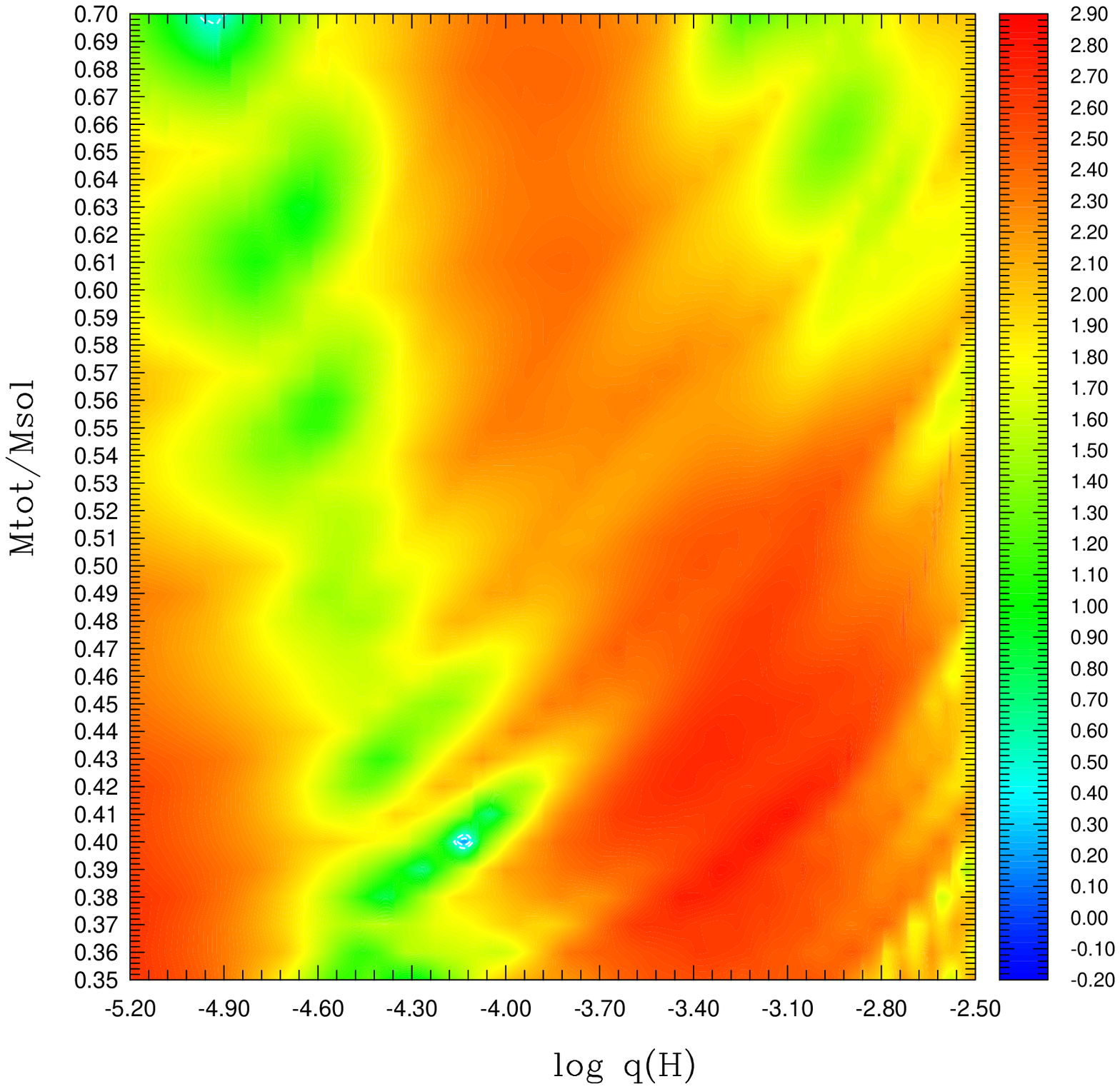}} \\
\end{tabular}
\caption{\it{Left Panel:} \rm slice of the $S^2$ function (in logarithmic units) along the $\log{g}-T_{\rm eff}$ plane at fixed parameters $M_{\ast}$ and $\log{q(\rm H)}$ set to their optimal values found for the best-fit solution Model 1 ($M_{\ast}$=0.40 $M_{\odot}$ and $\log{q(\rm H)}$=$-$4.11). The solid line rectangle represents our mean spectroscopic estimate of the atmospheric parameters with their uncertainties. Also indicated are the 1, 2 and 3 $\sigma$ limits on the $S^2$ minimum (white dashed contours). The errors on the asteroseismologically derived parameters are estimated from the semi-axes of the 1 $\sigma$ contour. \it{Right Panel}: \rm slice of the $S^2$ function (in logarithmic units) along the $M_{\ast}-\log{q(\rm H)}$ plane at fixed parameters $\log{g}$ and $T_{\rm eff}$ set to their optimal values found for the best-fit solution Model 1 ($\log{g}$=5.686 $M_{\odot}$ and $T_{\rm eff}$=31,940 K). The 1, 2 and 3 $\sigma$ limits on the $S^2$ minimum are once again indicated.}
\label{mod1chi2}
\end{figure*}

The observations of PG 0911+456 provide us with 7 harmonic oscillations to compare with theoretical predictions. This seems a reasonable number for attempting an asteroseismological analysis, since similar studies have been carried out on the basis of anywhere between 3 (EC 20117-4014, \citet{randall2006c}) and 13 (PG 0014+067, \citet{brassard2001}) periodicities. Nevertheless, the model finding algorithm uncovers several families of models that can account for the detected period spectrum similarly well, complicating the identification of a convincing "optimal" model. By carefully exploring parameter space under different assumptions as to the allowed degree indices of the modes observed we nevertheless isolate a promising solution.

\subsection{Search for the optimal model for PG 0911+456}

We started the search for the optimal model by keeping any a priory assumptions to a minimum. As usual, the pulsation calculations were effectuated for all modes with degree indices 0 $\leq \ell \leq$ 4 in a period range amply covering the observed oscillations (100$-$300 s). Higher degree modes are thought to have amplitudes too small for them to be detectable due to cancellation effects when integrating over the visible disk of the star. Indeed, the observability of modes with degrees higher than 2 was long debated, but the period density and distribution uncovered in certain EC 14026 stars (e.g. KPD 1930+2752, \citet{billeres2000} or PG 1325+101, \citet{charp2006}) forced the inclusion of higher degree modes. In the case of PG 0911+456 this is not strictly necessary as the pulsation periods observed \it{can} \rm be accounted for solely in terms of modes with $l\leq$ 2 (see below). However, we feel that it would be counterproductive to exclude potentially interesting models from the outset. 

Throughout the exploration of parameter space, we fixed the effective temperature to the mean spectroscopic value of 31,940 K, thereby decreasing the number of free parameters from 4 to 3. In our experience, this facilitates and speeds up the identification of potential "optimal" models. Given the weak temperature dependence of the pulsation periods, $T_{\rm eff}$ is in any case determined more accurately on the basis of spectroscopy than asteroseismology (the inverse holds true for $\log{g}$, which can be constrained far more tightly from asteroseismology by virtue of its strong impact on the period spectrum). And indeed, test calculations carried out at fixed temperatures at the limits of the spectroscopic uncertainties (employing $T_{\rm eff}$=32,200 K and $T_{\rm eff}$=31,800 K) produced the same families of optimal models, albeit with a slight ($\sim$ 0.01 $M_{\odot}$) shift in mass in each case. The compensation of an increase in temperature with an increase in mass is well understood in terms of parameter degeneracy and has been observed in the asteroseismic exploration of most other EC 14026 stars (e.g. PG 1219+534, \citet{charp2005a}). It implies that the two parameters cannot be optimised independently, obliging us to fix or at least constrain the temperature if the mass is to be determined to any accuracy. As for the remaining model input parameters, we sandwich the value suggested from spectroscopy for the surface gravity, searching the range 5.50 $\leq\log{g}\leq$ 5.80, while possible values of the stellar mass and the hydrogen envelope thickness are constrained from evolutionary models to 0.30 $\leq M_{\ast}/M_{\odot} \leq$ 0.70 and $-$5.20 $\leq \log{q(H)} \leq$ $-$2.00 (see \citet{charp2005a} for more details).   

Within the search domain specified, our algorithm identified several families of models that present good period matches. The majority of these were discarded upon closer inspection, either because of obvious inconsistencies with the spectroscopic values of $\log{g}$, or because the mode identification inferred was deemed inappropriate (e.g. the dominant pulsation was assigned $\ell$=3). However, two of the best-fit models (i.e. those that matched the observed periods with the lowest values of $S^2$) retained our attention as potential "optimal" models and are discussed in more detail in what follows. 

The first minimum in the goodness-of-fit merit function (at $S^2\sim$ 0.63) corresponds to a model with $T_{\rm eff}$ = 31,940 K (fixed from spectroscopy), $\log{g}$=5.686, $M_{\ast}$=0.40 $M_{\odot}$ and $\log{q(H)}$=$-$4.11 (hereafter referred to as Model 1). Figure \ref{mod1chi2} illustrates the complex behaviour of the $S^2$ hyper-surface in the vicinity of this solution: the left hand panel shows a slice along the $\log{g}-T_{\rm eff}$ plane with the mass and envelope thickness set to their optimum values, whereas the right hand panel displays a cut through $M_{\ast}-\log{q(\rm H)}$ space with the temperature and surface gravity fixed to the Model 1 solution. As indicated on the scale to the right of the main plots, areas shaded in dark blue indicate a relatively good match between the observed and theoretical periods, while those coloured in red represent comparatively bad fits. From the left hand panel we can see that the good-fit models are concentrated in two elongated $S^2$ "valleys", one centred around $\log{g}\sim$ 5.68 (Model 1), and the other around $\log{g}\sim$ 5.61. The second of these is quite clearly inconsistent with the mean spectroscopic estimate (indicated by the black rectangle) and can be rejected immediately. Model 1 on the other hand may lie well outside the spectroscopic box shown, however a closer look reveals that the surface gravity inferred is quite close to that derived from our MMT spectrum. Despite our conclusion that the latter is less accurate than the higher $\log{g}$ value suggested by both the Kitt Peak data and \citet{koen1999} (see Section 2.1), it seems foolhardy to reject this solution immediately.   

The second possible family of optimal models (with $S^2\sim$ 1.64) is centred at $T_{\rm eff}$=31,940 K (fixed from spectroscopy), $\log{g}$=5.777, $M_{\ast}$=0.39 and $\log{q(\rm H)}$=$-$4.78 (hereafter referred to as Model 2). Although the value of $S^2$ is more than a factor of 2 higher than that for Model 1, the period match is nevertheless perfectly acceptable at $<\Delta P/P>\sim$ 0.26 \% or, equivalently, an absolute dispersion of 0.41 s, and quite typical of the precision achieved in asteroseismological studies to date. Moreover, the value of the surface gravity determined is nearly identical to our most reliable spectroscopic estimate. Seeing as though both Model 1 and Model 2 predict all the theoretical modes matched with observed pulsations to be unstable, non-adiabatic theory does not further discrimination between the two. Instead, we turn to the mode identification inferred.

\begin{table}
\caption{Mode identification inferred from Model 1 and Model 2.}
\label{modeidentification}
\centering
\begin{tabular}{c c|c c|c c}
\hline\hline
& & \multicolumn{2}{c}{Model 1} & \multicolumn{2}{c}{Model 2} \\
\hline
Rank & $P_{obs}$ (s) & $\ell$ & $k$ & $\ell$ & $k$ \\
\hline\hline 
1 & 155.767 &  0 & 2 & 0 & 1 \\
2 & 165.687 &  2 & 2 & 0 & 0 \\
3 & 161.554 &  3 & 2 & 2 & 1 \\
4 & 192.551 &  2 & 1 & 2 & 0 \\
5 & 157.581 &  4 & 2 & 4 & 1 \\
6 & 168.784 &  1 & 2 & 4 & 0 \\
7 & 149.027 &  1 & 3 & 1 & 2 \\
\hline
\end{tabular}
\end{table}

\begin{figure}
\resizebox{\hsize}{!}{\includegraphics{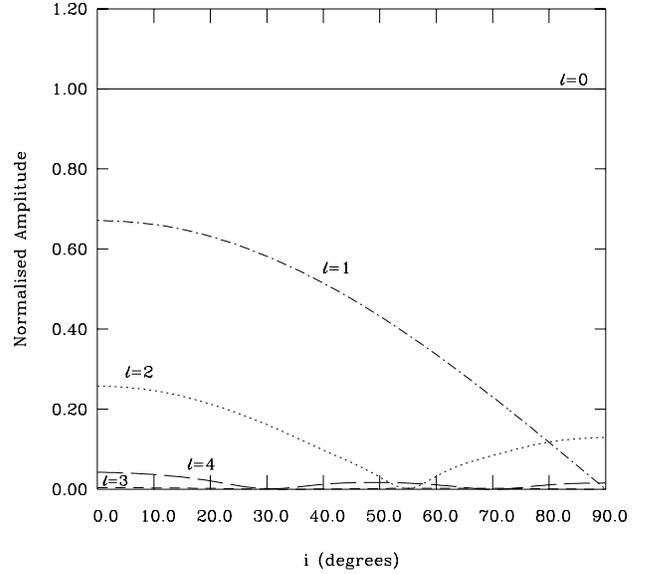}}
\caption{Variation of the disk-integrated pulsational amplitudes with inclination angle $i$ for $m$=0 modes with $\ell$=0 (continuous line), $\ell$=1 (dot-dashed), $\ell$=2 (dotted), $\ell$=3 (dashed - hardly visible due to the extremely low amplitudes) and $\ell$=4 (long-dashed) for an atmospheric model of PG 0911+456. The amplitudes were normalised with respect to the radial mode. Note that the \it{intrinsic} \rm amplitudes were  assumed to be the same for all modes.}
\label{amplitudes}
\end{figure}

Table \ref{modeidentification} lists the observed pulsations in order of decreasing amplitude rank, together with the resulting mode identification in terms of degree $\ell$ and radial order $k$ for the two solutions. In the case of Model 1, the degree indices assigned to the observed oscillations are not entirely convincing. While the dominant pulsation is associated with a radial mode as would be expected, both on the basis of visibility arguments and partial mode identification from multi-colour photometry for other EC 14026 stars (the main pulsations in both KPD 2109+4401 and Balloon 090100001 were shown to be radial, see \citet{randall2005} and \citet{fontaine2006} respectively), the amplitude rank 3 oscillation is identified with an $\ell$=3 mode. Detailed model atmosphere calculations have recently shown that such modes have negligible amplitudes and are in fact geometrically less visible than those with $\ell$=4 (\citet{randall2005}; see also Figure \ref{amplitudes}). Moreover, multi-colour photometry of pulsating subdwarf B stars has \it{never} \rm indicated the presence of $\ell$=3 modes at detectable amplitudes, and specifically excluded this possibility for some 27 oscillations observed in 7 different stars (see \citet{tremblay2006}, \citet{jeffery2005}). Note that these pulsations show a very different amplitude-wavelength behaviour to modes with $\ell$=0,1,2,4 and should therefore be readily identifiable if they are found. Modes with $\ell$=4 on the other hand have been inferred in a few cases, e.g. for KPD 2109+4401, HS 0039+4302 \citep{jeffery2004} and Balloon 090100001 (Brassard et al. 2007, in preparation). 

Model 2 seems rather more convincing in terms of the degree and radial orders inferred. For one, the amplitude hierarchy goes strictly with $\ell$: the dominant two modes are found to be radial, followed by two quadrupole modes, two pulsations with $\ell$=4 and, finally, a lone dipole mode. Secondly, with the exception of the weakest oscillation, the modes observed for each degree correspond to the fundamental mode and the first overtone, indicating \it{bands} \rm of instability rather than isolated frequencies excited to observable amplitudes. Although our pulsation calculations rely on linear theory and consequently cannot predict the amplitudes of modes, it is more intuitive to imagine that within a certain frequency range all modes should be excited to a comparable intrinsic amplitude. And, indeed, asteroseismological studies of other EC 14026 pulsators find that the majority of observed modes, particularly those with relatively high amplitudes, lie at the low-frequency end of the $p$-mode branch.

\begin{figure*}
\begin{tabular}{cc}
{\includegraphics[width=8.8cm]{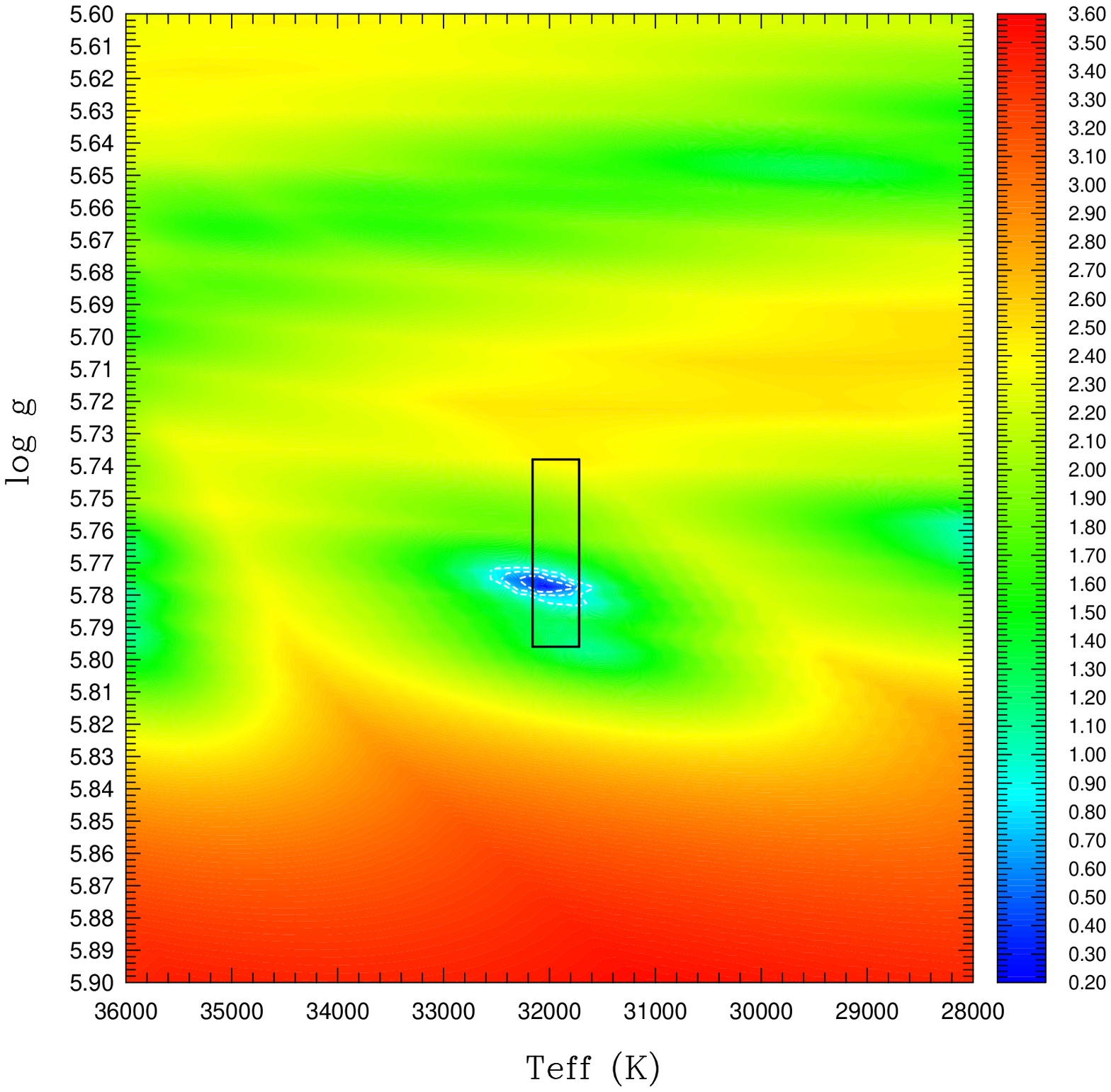}} & {\includegraphics[width=8.8cm]{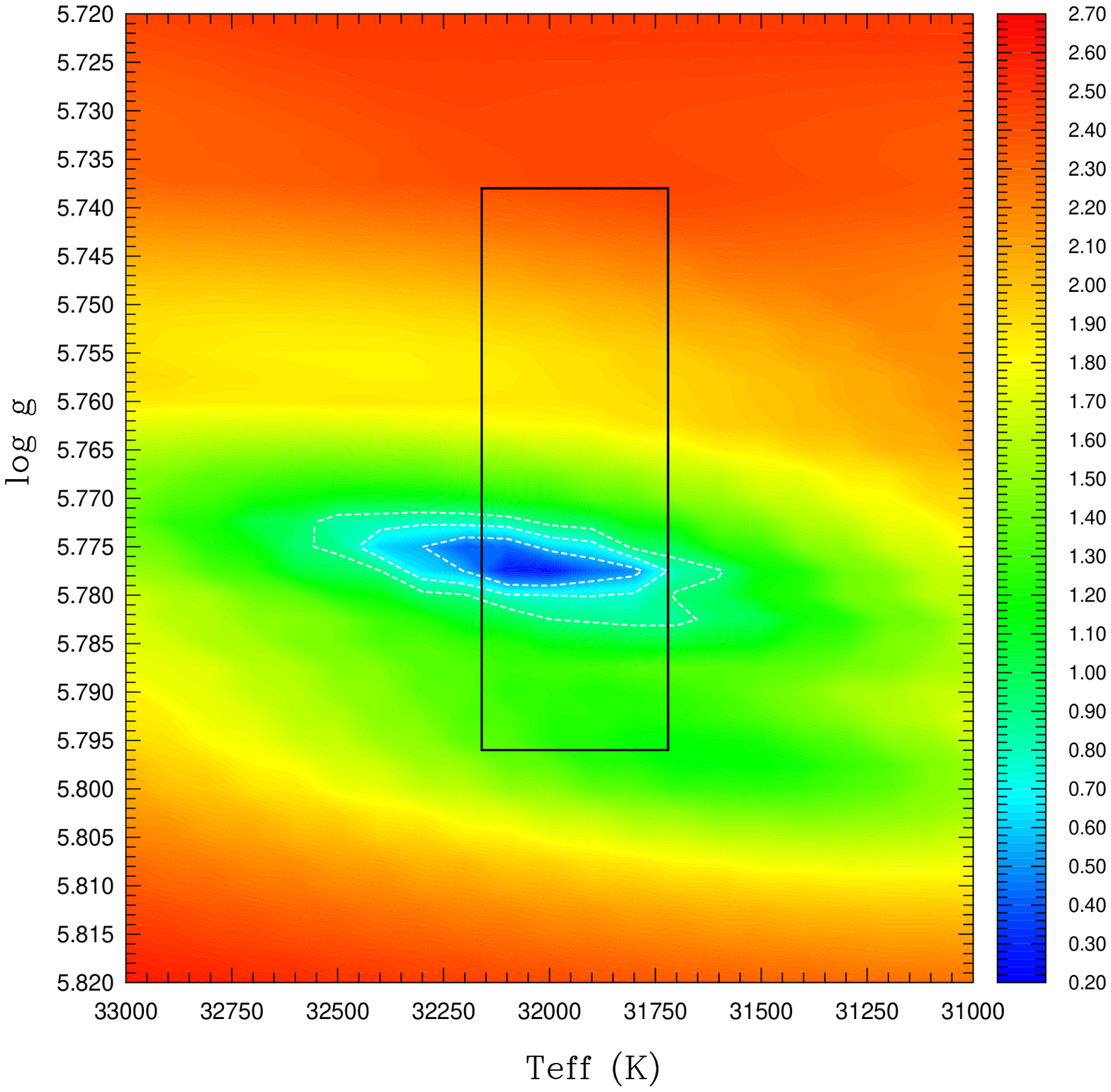}} \\
\end{tabular}
\caption{\it{Left Panel:} \rm slice of the $S^2$ function (in logarithmic units) along the $\log{g}-T_{\rm eff}$ plane at fixed parameters $M_{\ast}$ and $\log{q(\rm H)}$ set to their optimal values found for the best-fit solution ($M_{\ast}$=0.39 $M_{\odot}$ and $\log{q(\rm H)}$=$-$4.78). The solid line rectangle represents our mean spectroscopic estimate of the atmospheric parameters with their uncertainties. Also indicated are the 1, 2 and 3 $\sigma$ limits on the $S^2$ minimum (white dashed contours). The errors on the asteroseismologically derived parameters are estimated from the semi-axes of the 1 $\sigma$ contour. \it{Right Panel:} \rm slice of the $S^2$ function zoomed in on the minimum.}
\label{mod2loggteff}
\end{figure*}

\begin{figure*}
\begin{tabular}{cc}
{\includegraphics[width=8.8cm]{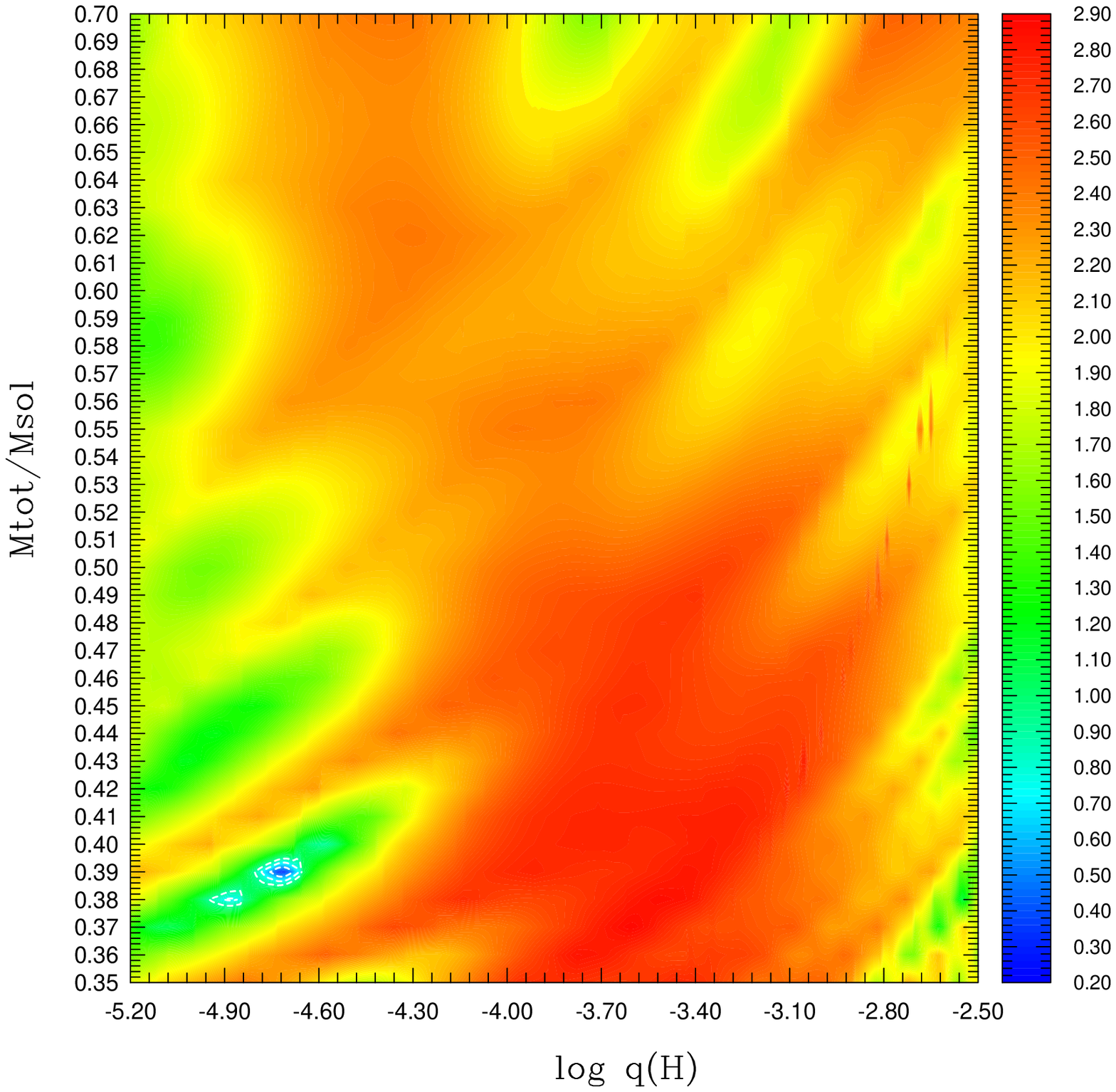}} & {\includegraphics[width=8.8cm]{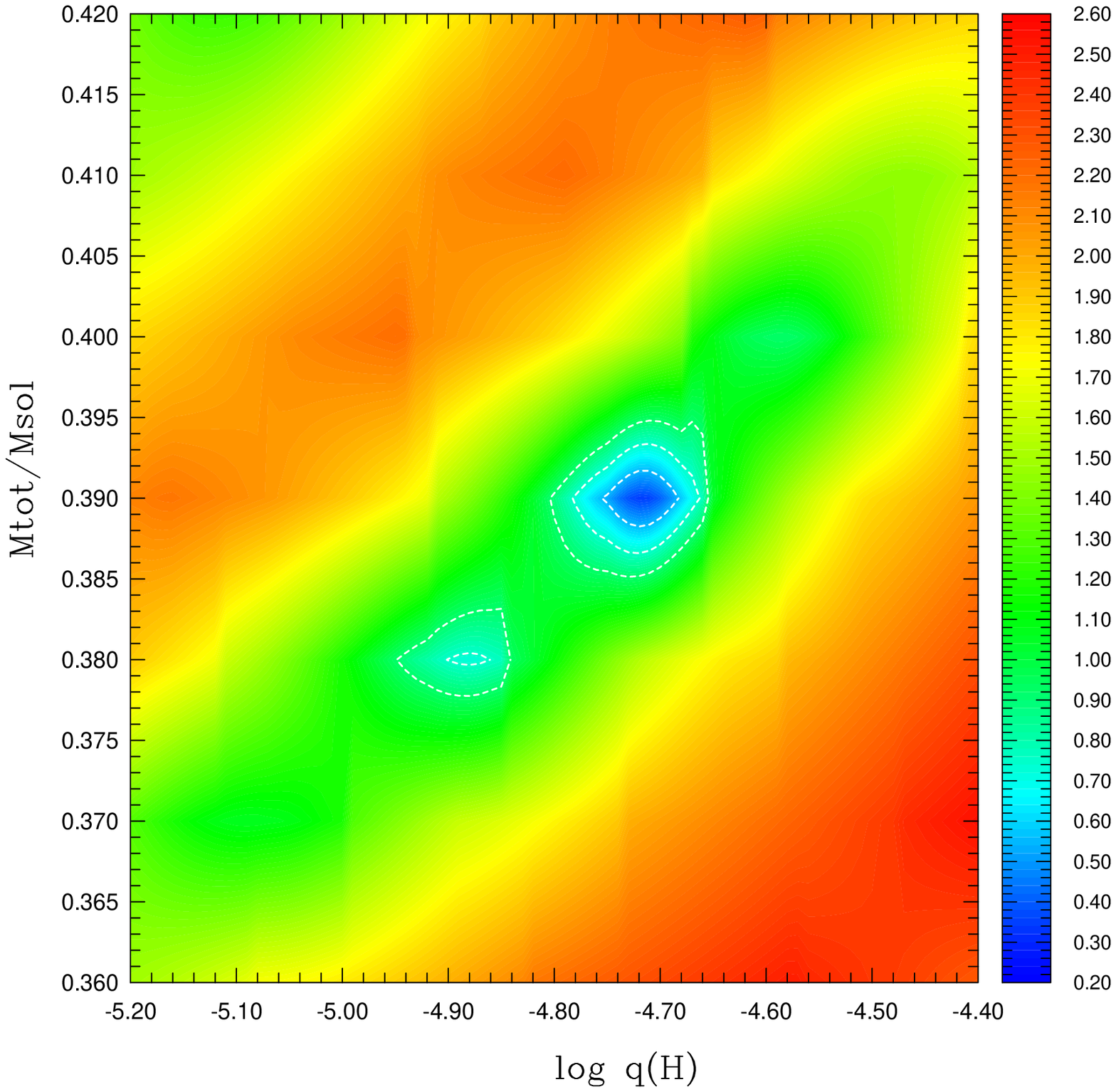}} \\
\end{tabular}
\caption{\it{Left Panel:} \rm slice of the $S^2$ function (in logarithmic units) along the $M_{\ast}-\log{q(\rm H)}$ plane at fixed parameters $\log{g}$ and $T_{\rm eff}$ set to their optimal values found for the best-fit solution ($\log{g}$=5.777 $M_{\odot}$ and $T_{\rm eff}$=31,940 K). Also indicated are the 1, 2 and 3 $\sigma$ limits on the $S^2$ minimum (white dashed contours). \it{Right Panel:} \rm slice of the $S^2$ function zoomed in on the minimum.}
\label{mod2Mlogqh}
\end{figure*}  

One potential shortcoming of both Model 1 and Model 2 is the conspicuous absence of dipole modes among the higher amplitude oscillations listed in Table \ref{modeidentification}. According to canonical wisdom, the amplitude at which a mode is observed should decrease rapidly with increasing degree index due to cancellation effects when integrating over the visible disk of the star. However, detailed calculations employing full model atmospheres (taking into account limb darkening, as well as radial, surface gravity and temperature variations over a pulsation cycle) show that this is not necessarily true. It was already shown by \citet{randall2005} that $\ell$=4 modes are more visible than those with $\ell$=3. Here, we go one step further and investigate the relative amplitude of modes with different degree indices as a function of the inclination angle $i$ (defined as the angle between the line-of-sight and the pulsation axis) employing the modified model atmosphere code for subdwarf B stars developed in Montr\'eal. This code incorporates several programs utilising the same bank of full model atmospheres and was used and described extensively by \citet{randall2005}. Given values for $\log{g}$ and $T_{\rm eff}$ as well as an inclination angle $i$, the wavelength bandpass of interest, and the characteristics of the oscillation mode, it outputs the emergent perturbed Eddington flux at user-specified points in time, and consequently enables the determination of relative pulsational amplitudes. In our calculations, we fixed the atmospheric parameters to those derived for PG 0911+456 from spectroscopy, set the bandpass to white light received at a representative observing site (in this case Kitt Peak National Observatory), and calculated the integrated amplitudes expected for low-degree (0 $\leq \ell \leq$ 4) modes for inclination angles from 0 to 90$^\circ$ assuming they have the same \it{intrinsic} \rm amplitudes. Note that the azimuthal order $m$ was set to 0 as appropriate for a slow rotator.  The results are illustrated in Figure \ref{amplitudes}, which shows the observable amplitudes expected for $\ell$=1,2,3 and 4 modes normalised relative to the radial mode (the amplitude of which is of course independent of the inclination) as a function of inclination $i$. It is quite striking that the canonical $\ell$=0,1,2,... amplitude hierarchy does not always hold true: in particular, dipole modes experience a rapid decrease in visibility with increasing inclination, and quadrupole modes have a higher integrated amplitude above $\sim$80$^\circ$, while being virtually undetectable at $\sim$55$^\circ$. As expected, $\ell$=3 modes have very low visibility regardless of the inclination angle. 

Returning to the mode identification inferred from our two models, we find that the absence of high-amplitude dipole modes could be explained by a very large inclination angle ($i\gtrsim$ 80$^\circ$). Moreover, the amplitude$-$degree index hierarchy inferred for Model 2 then fits in very nicely with the plot, that for Model 1 less so due to the high-amplitude $\ell$=3 mode identified. Of course, we are working under the assumption that the intrinsic oscillation amplitudes of the modes are similar, which in the absence of non-linear pulsation theory we have no means of verifying. Nevertheless, we believe Figure \ref{amplitudes} carries the important message that the canonical simplified $\ell$=0,1,2... amplitude hierarchy is not always valid, and that the visibility of a mode is strongly dependent on the inclination at which it is observed.

Having established that modes with $\ell$=4 are more likely to be visible than those with $\ell$=3, we conducted a second search for an optimal model specifically excluding the latter. The ranges considered for the input parameters were kept the same as during the first search. Perhaps unsurprisingly considering the quality of the fit, the global minimum was found to be identical with the Model 2 isolated during the first sweep of parameter space. Figures \ref{mod2loggteff} and \ref{mod2Mlogqh} show the corresponding slices of the $S^2$ function along the $\log{g}-T_{\rm eff}$ and the $M_{\ast}-\log{q(\rm H)}$ plane respectively, with the remaining parameters fixed to their optimal values. As already pointed out above, the value of $\log{g}$ inferred is in excellent agreement with the mean spectroscopic estimate. It is also apparent that there are \it{two} \rm shallow $S^2$ valleys along the effective temperature axis, one centered on Model 2, and the other around $\log{g}\sim$ 5.66, close to Model 1. This is not a coincidence; in fact, we demonstrate below that Model 1 and 2 represent analogous solutions, with a mode identification shifted by $k$=$-$1 from the lower to the higher gravity $S^2$ valley. Note that in an extension of this "mode jumping" phenomenon the third $S^2$ minimum at $\log{g}\sim$ 5.61 in Figure \ref{mod1chi2} is associated with a $k$=+1 shift from Model 1. This behaviour can be understood in terms of a relative uniformity of the period distribution for a given $\ell$ value and the strong period dependence on $\log{g}$. When the surface gravity of a model is slowly moved away from that of the optimal model, the theoretical period spectrum shifts compared to that observed, and the quality of the fit degrades at first. Once the change in $\log{g}$ becomes large enough, the $S^2$ function slowly decreases, converging towards the next solution, with a mode identification shifted by 1 in radial order. Luckily for asteroseismology, the low-order $p$-modes detected in EC 14026 stars are quite sensitive to the internal composition of the star, and the modes are normally not spaced too uniformly in $k$, enabling a discrimination between different solutions on the basis of $S^2$ in conjunction with the spectroscopic constraints on $\log{g}$ and $T_{\rm eff}$ (see Figure \ref{fit}). In the high-order asymptotic regime the story is very different, and this is in fact one of the main problems when quantitatively interpreting slowly pulsating subdwarf B stars (see \citet{randall2006b}). 

In order to assess the significance and relative importance of Model 2, we carried out several additional searches in 3-dimensional parameter space under artificially imposed constraints. First, we limited the acceptable degree indices to $\ell\leq$ 2 to determine whether the observed period distribution could be recreated solely in terms of the traditionally expected modes. While we found a physically viable model, the quality of the fit ($S^2\sim$ 38) was so severely degraded compared to the previous models that it could not be seriously considered as an optimal model. Next, we forced the degree index identification inferred from Model 1 and 2 onto the higher and lower gravity valleys respectively, and were able to recover local minima very close to the other model, albeit at a much degraded quality-of-fit. In both cases, the mode identification was shifted in $k$, Model 2 starting at $k$=0 and Model 1 at $k$=1 in a convincing illustration of "mode jumping". Finally, we imposed the canonically expected amplitude$-$degree hierarchy on mode identification during the optimal model search, limiting the higher amplitude modes to $\ell\leq$ 2, while additionally allowing $\ell$=4 for the two lowest amplitude modes. From this we found a model very similar to Model 2, at a degraded but still acceptable quality-of-fit ($S^2\sim$ 9.8). We believe that this confirms the fundamental robustness of Model 2, and thus adopt it as the optimal model for PG 0911+456.

\subsection{Period fit for the optimal model}

\begin{figure}
\resizebox{\hsize}{!}{\includegraphics{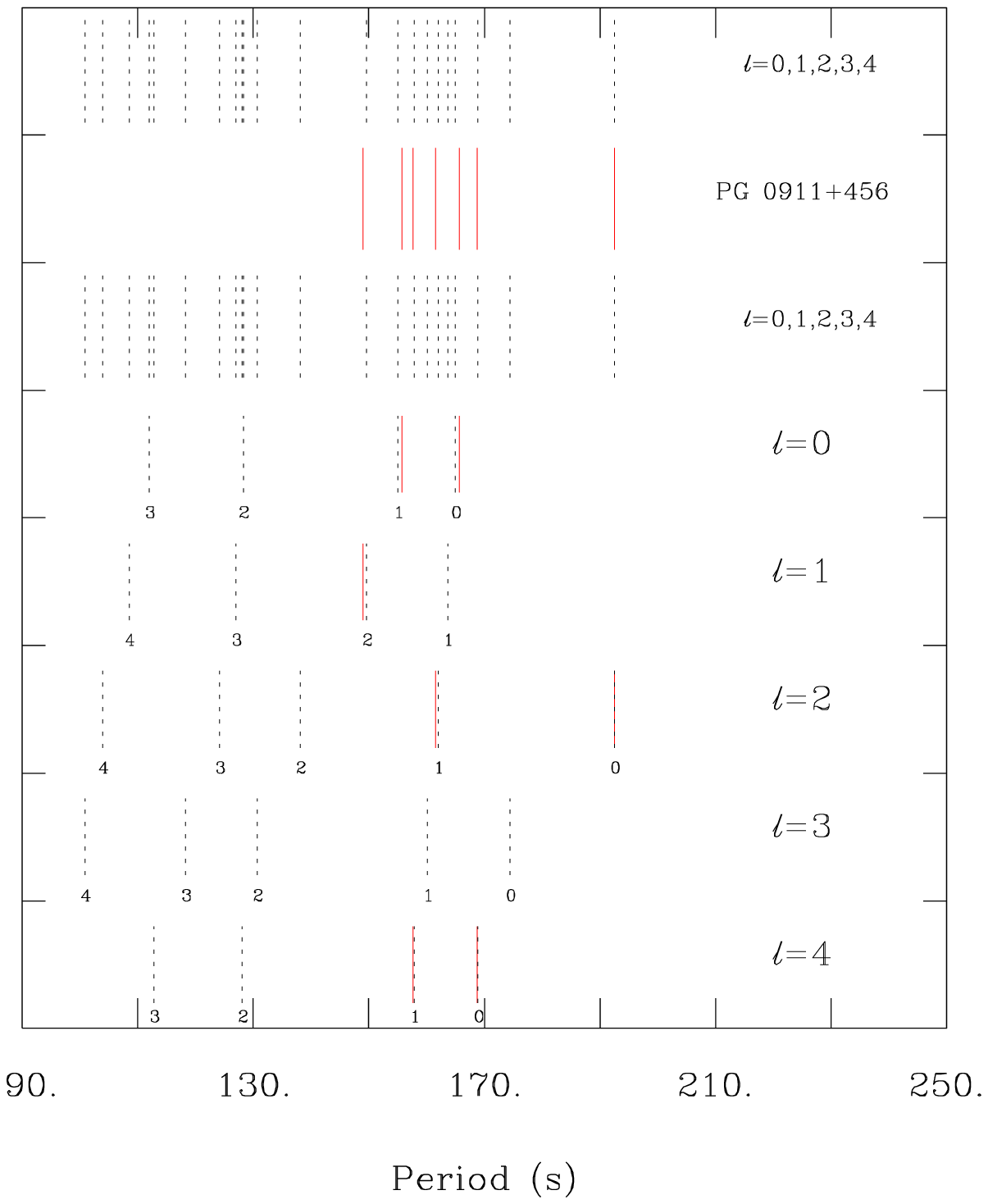}}
\caption{Comparison of the observed period spectrum of PG 0911+456 (continuous line segments) with the theoretical pulsation spectrum of the optimal model (dotted line segments) in the 100$-$200 s period range for degree indices $\ell$=0,1,2,3,4. Note that all the theoretical modes plotted are predicted to be excited and correspond to low-order $p$-modes. The radial order $k$ is indicated below each segment.}
\label{fit}
\end{figure}

The optimal model isolated for PG 0911+456 provides an excellent match to the 7 independent periods clearly identified in this star. Details on the period fit and mode identification are given in Table \ref{periodfit} and are graphically represented in Figure \ref{fit}. As briefly mentioned earlier, the quality of the period fit at $\Delta P/P\sim$ 0.26 \% is typical of that achieved for asteroseismological analyses of other EC 14026 stars; the dispersion is however still an order of magnitude higher than the accuracy to which the observed periods can be measured. This has been the case for most EC 14026 stars analysed to date and is attributed to remaining imperfections in the equilibrium models. It is of course one of the goals of our asteroseismic studies to fine-tune these models, a quest that we are currently actively working on. Nevertheless, we believe that the fact that we were able to isolate an optimal model that a) is in accordance with spectroscopy, b) can accurately reproduce the observed period spectrum and c) predicts all observed modes to be unstable confirms the basic validity of our sdB star models. 

\begin{table*}
\caption{Period fit for the optimal model. We list the mode identification of the observed periods together with the theoretical periods, the stability coefficient $\sigma_I$, the kinetic energy $\log{E}$ and the rotation coefficient $C_{kl}$. Also indicated are the relative dispersion of the period fit and the amplitude rank of the observed periodicity.}
\label{periodfit}
\begin{center}
{\small \begin{tabular}{c c c c c c c c c}\hline
{}&{}&{$P_{obs}$}&{$P_{th}$}&{$\sigma_{I}$}&{log
\textit{E}}&{$C_{kl}$}&{$\Delta P/P$}&{Comments}\\
{$l$}&{$k$}&{(s)}&{(s)}&{(rad/s)}&{(erg)}& {} &(\%)&{}\\
\hline 0&3&...&112.010&$-1.070\times
10^{-5}$&40.781&...&...&\\
0&2&...&128.259&$-1.233\times
10^{-5}$&40.751&...&...&\\
0&1&155.767&155.053&$-1.048\times
10^{-6}$&41.651&...&$+0.4585$&$\sharp$1\\
0&0&165.687&164.995&$-7.363\times
10^{-7}$&41.741&...&$+0.4175$&$\sharp$2\\
&&&&&&&&\\
1&4&...&108.498&$-1.404\times
10^{-5}$&40.632&0.01728&...&\\
1&3&...&127.004&$-1.357\times
10^{-5}$&40.710&0.01367&...&\\
1&2&149.027&149.644&$-9.475\times
10^{-7}$&41.720&0.03214&$-0.4140$&$\sharp$7\\
1&1&...&163.618&$-1.080\times
10^{-6}$&41.588&0.01923&...&\\
 &&&&&&&&\\
 2&4&...&103.956&$-2.518\times
10^{-5}$&40.323&0.02767&...&\\
2&3&...&124.148&$-1.408\times
10^{-5}$&40.690&0.03422&...&\\
2&2&...&138.108&$-2.249\times
10^{-6}$&41.409&0.08731&...&\\
2&1&161.554&161.986&$-1.390\times
10^{-6}$&41.487&0.02621&$-0.2673$&$\sharp$3\\
2&0&192.551&192.519&$-8.212\times
10^{-10}$&44.354&0.36050&$+0.0168$&$\sharp$4\\
&&&&&&&&\\
3&4&...&100.866&$-3.466\times
10^{-5}$&40.130&0.02603&...&\\
3&3&...&118.293&$-1.096\times
10^{-5}$&40.785&0.07479&...&\\
3&2&...&130.678&$-7.973\times
10^{-6}$&40.913&0.05462&...&\\
3&1&...&160.138&$-1.538\times
10^{-6}$&41.445&0.04046&...&\\
3&0&...&174.449&$-3.289\times
10^{-8}$&42.952&0.17867&...&\\
&&&&&&&&\\
4&3&...&112.770&$-1.088\times
10^{-5}$&40.767&0.06969&...&\\
4&2&...&128.086&$-1.149\times
10^{-5}$&40.764&0.03321&...&\\
4&1&157.581&157.851&$-1.542\times
10^{-6}$&41.444&0.05935&$-0.1715$&$\sharp$5\\
4&0&168.784&168.916&$-1.644\times
10^{-7}$&42.302&0.11349&$-0.0780$&$\sharp$6\\
4&1&...&238.583&$-3.504\times10^{-12}$&46.073&$-0.01391$&...&\\
 \hline
\end{tabular}}
\end{center}
\end{table*}

From Figure \ref{fit} it is apparent that the observed periods are exclusively identified with modes clustered at the high-period end of the theoretical $p$-mode branch. This behaviour is reminiscent of that uncovered for other EC 14026 pulsators such as PG 0014+067 \citep{brassard2001} and PG 1219+534 \citep{charp2005a}, and seems to support the idea that energy powering up the pulsations may be preferentially distributed among the modes of low radial order. These modes would then reach observable amplitudes more easily (depending of course on the geometric mode visibility) and would therefore be detected in the light curves more readily. It should be noted however, that some EC 14026 pulsators (e.g. PG 1325+101, \citet{charp2006}) also show isolated modes at lower periods, with apparent gaps in the radial order distribution. These cannot be fully understood without non-linear pulsation theory, but it seems probable that some of the "missing" modes will eventually be detected. In the past, observations at higher S/N have $\it{always}$ \rm revealed extra oscillations compared to noisier data, supporting the hypothesis that sdB stars often excite many more modes than are uncovered, albeit at lower observable amplitudes. Apart from geometric visibility, the observability of oscillations also depends on the intrinsic amplitudes, which depend on the exact physical conditions inside the star integrated over the propagation region of each mode. There is also some evidence of oscillation power being shifted between modes over time, measurable from the relative amplitude changes of the observed pulsation periods (insofar as beating between closely spaced modes can be excluded). PG 0911+456 is an example of this: in the 9 years between the observations of \citet{koen1999} and our own, the amplitude of the dominant peak compared to the second and third periodicity has increased by a factor of nearly 3, suggesting a transfer of power into the preferential driving of the highest amplitude mode. It remains to be investigated whether this is a long-term trend and, indeed, whether the amplitude variations reported for other sdB pulsators follow any discernable pattern. This will be an interesting incentive for future observational projects.

\subsection{Structural parameters of PG 0911+456}

The optimal model identified leads to a natural determination of the three variable input parameters for PG 0911+456, these being the surface gravity $\log{g}$, the stellar mass $M_{\ast}$, and the mass of the hydrogen-rich envelope through the quantity $\log{q(\rm H)}$. In addition, we have the effective temperature $T_{\rm eff}$ from spectroscopy. On the basis of these quantities we can derive a set of secondary parameters: the stellar radius $R$ (as a function of $M_{\ast}$ and $g$), the luminosity $L$ (as a function of $T_{\rm eff}$ and $R$), the absolute magnitude $M_V$ (as a function of $g$, $T_{\rm eff}$ and $M_{\ast}$ in conjunction with the use of detailed model atmospheres) and the distance from Earth $d$ (as a function of apparent magnitude and $M_V$). Given that the Str\"omgen magnitude of $b\sim$14.6 estimated by \citet{koen1999} is associated with large and not readily quantifiable errors, we followed the suggestion of the referee and employed the SDSS values instead, these being $u'$=14.073$\pm$0.005, $g'$=14.458$\pm$0.006 and $r'$=14.885$\pm$0.005 (accessible through Vizier@CDS). Using the calibration equations given in \citet{smith2002} we derive a Johnson magnitude of $V$=14.663, associated with a formal error of 0.007 mags. However, the latter is likely to be severely underestimated as the transformation from the SDSS to the Johnson-Cousins colour system introduces additional uncertainties. Comparing the $V$ and $B-V$ values of 4 sdB stars calibrated on the basis of the SDSS values to those given by \citet{allard1994} reveals typical differences in the magnitudes on the order of 0.03$-$0.05 mags. For the relatively faint PG 0911+456 we therefore adopt $V$=14.663$\pm$0.050, and use this in the computation of the distance. Tentative limits could also be set on the rotation period $P_{rot}$ and the equatorial rotation velocity $V_{eq}$ (as a function of $R$ and $P_{rot}$) due to the absence of fine frequency structure in the observed period spectrum, assuming that any multiplet components would have amplitudes above the detection limit. The values and limits for the parameters derived are summarised in Table \ref{modelparams}. Uncertainties on the three primary asteroseismological quantities $\log{g}$, $M_{\ast}$ and $\log{q\rm(H)}$ were estimated from the semi-axes of the 1 $\sigma$ contours shown in Figures \ref{mod2loggteff} and \ref{mod2Mlogqh}. These were calculated according to the recipe described in detail by \citet{brassard2001} and \citet{charp2005a}.

\begin{table}
\caption{Structural parameters of PG 0911+456.}
\label{modelparams}
\centering
\begin{tabular}{lrclc}
\hline\hline
 Quantity & \multicolumn{3}{c}{Estimated Value} & \tabularnewline
\hline\hline
 $T_{\rm eff}$ (K)$^{\dag}$ & $31940$ & $\pm$ & $220$ & ($0.69 \%$) \tabularnewline
 $\log g$ & $5.777$ & $\pm$ & $0.002$ & ($0.03 \%$) \tabularnewline
 $M_*/M_{\odot}$ & $0.39$ & $\pm$ & $0.01$ & ($1.55 \%$) \tabularnewline
 $\log (M_{\rm env}/M_*)$ & $-4.69$ & $\pm$ & $0.07$ & ($1.49 \%$) \tabularnewline
 & \tabularnewline
 $R/R_{\odot}$ ($M_*$, $g$) & $0.134$ & $\pm$ & $0.002$ & ($1.00 \%$) \tabularnewline
 $L/L_{\odot}$ ($T_{\rm eff}$, $R$) & $16.8$ & $\pm$ & $1.0$ & ($4.76 \%$) \tabularnewline
 $M_V$ ($g$, $T_{\rm eff}$, $M_*$) & $4.82$ & $\pm$ & $0.04$ & ($0.83 \%$) \tabularnewline
  $d$ ($V$, $M_V$) (pc) & $930.3$ & $\pm$ & $27.4$ & ($2.9 \%$) \tabularnewline
 $P_{\rm rot}$ (days) & & $\geq$ 68 & & ... \tabularnewline
 $V_{\rm eq}$ ($P_{\rm rot}$, $R$) (km/s) & & $\leq$ 0.1 & & ... \tabularnewline
\hline
{\footnotesize $^{\dag}$ From spectroscopy}
\end{tabular}
\end{table}

\begin{figure}
\resizebox{\hsize}{!}{\includegraphics{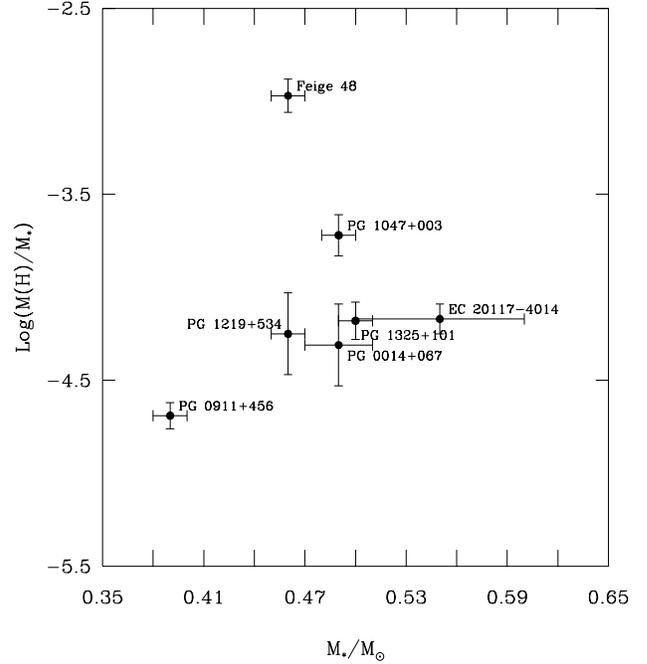}}
\caption{A graphical illustration of the total masses and relative hydrogen envelope masses determined for the 7 rapidly pulsating subdwarf B stars so far submitted to asteroseismology (for references see the Introduction).}
\label{masses}
\end{figure}

Comparing the structural parameters inferred for PG 0911+456 to those determined for other EC 14026 stars from previous asteroseismological analyses, we find that both the total mass and the envelope mass are lower than average (see Figure \ref{masses}). In fact, the total mass is the smallest ever measured for an sdB star and lies well below the canonical model value of $\sim$ 0.47 $M_{\odot}$, while still falling into the low-mass tail of the distribution expected \citep{dorman1993,han2002,han2003}. Assuming that PG 0911+456 is indeed a single star, the mass derived lies just below the expected range predicted from the white dwarf merger scenario proposed by \citet{han2002}; however, the authors specifically state that their values constitute upper limits since they assume that $\it{all}$ \rm the mass contained in the two merged helium white dwarfs goes into the subdwarf. In this context, the authors suggested searching for circumstellar matter around sdB's thought to be single stars. A positive detection would lower the minimum mass expected for products of the WD channel and perhaps explain the small mass determined for PG 0911+456\footnote{Note that the asteroseismic mass determination is robust. We carried out additional searches in parameter space forcing the mass to the canonical value of 0.47 $M_{\odot}$ and found that the resulting best fit solution presented an unacceptably bad period match with a dispersion of $<\Delta P/P>$=1.5 \% or $\Delta P$=2.5 s (compared to $<\Delta P/P>$=0.26 \% for the optimal model). Moreover, the mode identification inferred was less than convincing, the only radial mode being assigned to one of the lowest amplitude pulsations. The period fit was improved only marginally by varying the mass in the 0.46$-$0.48 $M_{\odot}$ range.}. As for the hydrogen shell mass, our results indicate that it is slightly smaller than that of most EC 14026 stars, even when considered relative to the low total stellar mass (see Figure \ref{masses}). This is precisely what would be expected if this star did indeed form via the WD merger scenario.

\section{Conclusion}

We obtained 57 hours of broad-band time-series photometry as well as high S/N low- and medium resolution time-averaged spectroscopy for the EC 14026 pulsator PG 0911+456. Our observations led to refined estimates of the star's atmospheric parameters and the detection of 7 independent harmonic oscillations, 4 more than were known previously. There was no sign of frequency splitting over the 68-day period during which the photometry was obtained, indicating a slow rotation rate. Fixing the effective temperature to the spectroscopic value and conducting an asteroseismic search in 3-dimensional $\log{g}-M_{\ast}-\log{q(\rm H)}$ parameter space enabled the identification of several families of models that could reproduce the observed periods to within less than 1 \%. While some of these were rejected from the outset due to obvious inconsistencies with the spectroscopic estimate of $\log{g}$ or implausible associated mode identifications, we retained two promising solutions for closer inspection. Unlike in some previous asteroseismological studies, it was not immediately obvious which model was to be preferred on the basis of the structural parameters alone; instead we used the inferred mode identification, in particular the degree index $\ell$, to discriminate between the two. The main difference  between the solutions was that one identified a relatively high amplitude peak with an $\ell$=3 mode, while the other required only modes with $\ell$=0,1,2 and 4. Since detailed computations reveal $\ell$=3 modes to have extremely small disk-integrated amplitudes that would most likely not be detectable, we favoured the latter and adopted it as the optimal model. 

The inferred structural parameters for PG 0911+456 include the total stellar mass and the thickness of the hydrogen-rich shell, two quantities that can normally not be derived using other means but are invaluable for a detailed understanding of subdwarf B stars' evolutionary history. The total mass determined is smaller than that found for any EC 14026 star to date and places our target at the low-mass end of the predicted distribution. Similarly, the hydrogen envelope is measured to be thinner than that of most sdB's studied so far. If PG 0911+456 is confirmed to be a single star, as is suspected from its slow rotation, negligible radial velocity variation, and absence of a companion's spectroscopic or near-IR photometric signature, it may be the product of a WD merger according to the evolutionary channels proposed by \citet{han2002,han2003}. In this case, we would indeed expect the hydrogen envelope mass to be smaller than for an sdB having undergone a CE or RLOF phase. The low total mass derived would tend to support a non-canonical evolutionary history, even if it lies slightly below the mass distribution predicted from a WD merger.   
    
An obvious follow-up study for PG 0911+456 is to verify the asteroseismological solution found on the basis of additional observations. These could aim for a higher S/N level, thus enabling the detection of further pulsations and strengthening the constraints on the asteroseismic model. One of the main challenges we faced in the search of parameter space was the relatively large number of models that could account for the oscillations observed quite accurately; it would be very instructive to see whether any newly found periods can also be fit by the optimal model isolated. Observations containing wavelength-dependent information from which modes may be partially identified are another option. Following recent theoretical investigations into the amplitude-wavelength dependence of a mode on its degree $\ell$ (\citet{ramachandran2004}, \citet{randall2005}), there has been a surge in observational efforts to obtain multi-colour photometry of EC 14026 stars, most notably using the 3-channel CCD ULTRACAM (e.g. \citet{jeffery2005}) and the Mont4kccd predecessor, LaPoune I (e.g. \citet{fontaine2006}). While discriminating between low-degree modes with $\ell$=0,1,2 has proved extremely challenging, $\ell$=4 modes exhibit a more clearly distinguishable amplitude-wavelength behaviour (see e.g. Figure 26 of \citet{randall2005}). Given the necessary data, it should therefore be possible to confirm the identification of the two $\ell$=4 modes inferred for PG 0911+456 and thus verify the structural parameters computed. 

The work presented here constitutes the 7th detailed asteroseismological analysis of a rapidly pulsating subdwarf B star. While we estimate that the structural parameters of around 20 targets are required to start detailed comparisons with evolutionary theory, first tentative efforts in this direction look promising. Nevertheless, it is clear that there is still ample room for improvement on the modelling front. Firstly, the fact that the dispersion between the observed and theoretical periods of the optimal model is generally an order of magnitude higher than the measurement accuracy indicates remaining shortcomings in the models. We are currently working on full evolutionary (rather than envelope) "third-generation" models to address this problem. The reliability of the "optimal" models identified during the search of parameter space is another issue. Although we do apply cross-checks such as compatibility with non-adiabatic theory and spectroscopic values of the atmospheric parameters, the latter (especially $T_{\rm eff}$) must often be used to discriminate between, or constrain, regions of minimum $S^2$ and can no longer be employed as independent estimates. Moreover, the period ranges of unstable oscillations computed from our non-adiabatic pulsation code are sensitive mostly to $T_{\rm eff}$ and $\log{g}$ and are largely independent of the model mass and envelope thickness. It is therefore vital that additional checks are carried out with regard to the robustness of the "forward" approach if the structural parameters inferred from asteroseismology are to be compared with evolutionary predictions in a quantitative manner. The most obvious way of doing this is by detecting more frequencies from higher S/N observations or constraining the identification of the degree $\ell$ of individual modes from multi-wavelength time-series data. Such efforts are ongoing, and will likely prove invaluable for the future of sdB star asteroseismology.

\begin{acknowledgements}
We would like to acknowledge Steward Observatory and the ITL for all their efforts in designing and building the Mont4k/La Poune II instrument. E.M. Green especially thanks Bill Peters for his usual invaluable assistance with scripting, and all the Mt. Bigelow staff for everything they do to make the observations better. We are also grateful to the Canada Foundation for Innovation equipment grant allocated to G. Fontaine through the Canada Research Chair. Finally, we are indebted to the referee Uli Heber, whose detailed comments and suggestions helped us to improve the work presented here. 
\end{acknowledgements}

\bibliographystyle{aa}
\bibliography{8433}

\end{document}